\documentclass[useAMSD,usenatbib]{mn2e}

\usepackage{aas_macros}
\usepackage{amsmath}
\usepackage{amssymb}
\usepackage{booktabs}
\usepackage{color}
\usepackage{relsize}
\usepackage{capt-of}
\usepackage[english]{babel}
\usepackage{epsfig}
\usepackage{float}
\usepackage[T1]{fontenc}
\usepackage{graphicx}
\usepackage[utf8]{inputenc}
\usepackage{mathrsfs}
\usepackage{subfig}
\usepackage{times}
\usepackage{hyperref}

\hypersetup{pagebackref=true, colorlinks=true,  linkcolor=red, citecolor=blue, urlcolor=blue, bookmarks=false}

\title[Mapping AGN accretion in the SFR--M$_{*}$ plane]{Mapping the average AGN accretion rate in the SFR--M$_{*}$ plane for \textit{Herschel}
\thanks{{\textit{Herschel} is an ESA space observatory with science instruments provided by European-led Principal Investigator consortia and with important participation from NASA.} } selected galaxies at 0$<$z$\leq$2.5
}

\author[I.~Delvecchio et al.]
{I.~Delvecchio$^{1,2}$\thanks{E-mail: ivan.delvecchio@unibo.it},
D.~Lutz$^2$, 
S.~Berta$^2$, 
D.~J.~Rosario$^2$, 
G.~Zamorani$^3$, 
F.~Pozzi$^1$,
\newauthor 
C.~Gruppioni$^3$,
C.~Vignali$^1$,
M.~Brusa$^{1,2,3}$,
A.~Cimatti$^1$,
D.~L.~Clements$^4$,
A.~Cooray$^{5,6}$,
\newauthor 
D.~Farrah$^7$,
G.~Lanzuisi$^{1,3}$,
S.~Oliver$^8$,
G.~Rodighiero$^{9}$,
P.~Santini$^{10}$,
and M.~Symeonidis$^{9,11}$
\\ \\
$^{1}$ Dipartimento di Fisica e Astronomia, Universit\`a di Bologna, via Ranzani 1, I-40127 Bologna, Italy. \\
$^{2}$ Max-Planck-Institut f\"ur extraterrestrische Physik, Giessenbachstrasse, 85748 Garching, Germany. \\
$^{3}$ INAF - Osservatorio Astronomico di Bologna, via Ranzani 1, I-40127 Bologna, Italy.\\
$^{4}$ Physics Department, Blackett Lab, Imperial College, Prince Consort Road, London SW7 2AZ, UK. \\
$^{5}$ Department of Physics \& Astronomy, University of California, Irvine, CA 92697, USA. \\
$^{6}$ California Institute of Technology, 1200 E. California Blvd., Pasadena, CA 91125, USA. \\
$^{7}$ Department of Physics, Virginia Tech, VA 24061, USA. \\
$^{8}$ Astronomy Centre, Dept. of Physics \& Astronomy, University of Sussex, Brighton BN1 9QH, UK. \\
$^{9}$ Dipartimento di Fisica e Astronomia ``G. Galilei'', Universit\`a di Padova, Vicolo dell’Osservatorio 3, I-35122, Italy. \\
$^{10}$ INAF - Osservatorio Astronomico di Roma, Via Frascati 33, I00040, Monte Porzio Catone, Italy. \\
$^{11}$ Mullard Space Science Laboratory, University College London, Holmbury St. Mary, Dorking, Surrey RH5 6NT, UK.\\
}

\begin{document}

\date{ Accepted 2015 January 28.  Received 2015 January 26; in original form 2014 November 13}

\pagerange{\pageref{firstpage}--\pageref{lastpage}} \pubyear{2015}

\maketitle

\label{firstpage}

\begin{abstract}
We study the relation of AGN accretion, star formation rate (SFR), and stellar mass (M$_*$) using a sample of $\approx$ 8600 star-forming galaxies up to z=2.5 selected with \textit{Herschel} imaging in the GOODS and COSMOS fields. For each of them we derive SFR and M$_*$, both corrected, when necessary, for emission from an active galactic nucleus (AGN), through the decomposition of their spectral energy distributions (SEDs). About 10 per cent of the sample are detected individually in \textit{Chandra} observations of the fields. For the rest of the sample we stack the X-ray maps to get average X-ray properties. After subtracting the X-ray luminosity expected from star formation and correcting for nuclear obscuration, we derive the average AGN accretion rate for both detected sources and stacks, as a function of M$_{*}$, SFR and redshift. The average accretion rate correlates with SFR and with M$_*$. The dependence on SFR becomes progressively more significant at z$>$0.8. This may suggest that SFR is the 
original driver of these correlations. We find that average AGN accretion and star formation increase in a similar fashion with offset from the star-forming ``main-sequence''. Our interpretation is that accretion onto the central black hole and star formation broadly trace each other, irrespective of whether the galaxy is evolving steadily on the main-sequence or bursting.
\end{abstract}


\section{Introduction} \label{intro}

A causal connection between super massive black hole (SMBH) and galaxy growth has been suggested by a number of studies, based on empirical correlations between black hole mass and integrated galaxy properties: galaxy bulge M$_*$, velocity dispersion (e.g. \citealt{Magorrian+98}; \citealt{Gebhardt+00}; \citealt{Ferrarese+02}; \citealt{Gultekin+09}). In addition, the cosmic star formation history and the black hole accretion history follow parallel evolutionary paths, peaking at z$\simeq$2 and declining towards the local Universe (\citealt{Boyle&Terlevich98}; \citealt{Shankar+09}; \citealt{Madau&Dickinson14}). 

Despite the mutual dependence on a common cold gas supply, such connections are not trivial given the vastly different spatial scales at which star formation (many kpc) and SMBH accretion (sub-pc) typically operate. Different scenarios have been proposed to justify the necessary loss of gaseous angular momentum, such as nuclear bars, minor and major merger events (e.g. \citealt{Garcia-Burillo+05}). However, the detailed mechanisms responsible for triggering black hole accretion and star formation are still poorly understood (e.g. see comprehensive review by \citealt{Alexander+12}).

Recent studies have highlighted a two-fold galaxy evolutionary scheme. More than 95 per cent of star-forming galaxies follow a reasonably tight relation frequently called the ``star-formation main-sequence'' (MS, scatter is about 0.2--0.4 dex) between SFR and M$_{*}$, from the local Universe up to z$\sim$3 (\citealt{Noeske+07}; \citealt{Daddi+07}; \citealt{Magdis+10}; \citealt{Elbaz+11}; \citealt{Whitaker+12}; \citealt{Schreiber+14}; \citealt{Speagle+14}). This trend is currently thought to reflect a large duty cycle of steady star formation in galaxies, fueled by a continuous gas inflow (\citealt{Dekel+09}; \citealt{Ciotti+10}). The most massive galaxies have larger gas reservoirs (\citealt{Tacconi+13}) and thus higher SFR. However, there are a few ($<$5 per cent) outliers in the SFR--M$_{*}$ plane with $>4$ times larger specific SFR (sSFR\footnote{sSFR is defined as the ratio between SFR and M$_*$.}). These off-sequence ``starbursts'' play a minor role in the cosmic star formation history (\citealt{
Rodighiero+11}) and show disturbed morphologies, probably due to galaxy interactions or gas-rich major mergers (e.g. \citealt{Hopkins&Hernquist09}; \citealt{Veilleux+09}). This distinction is supported by several studies, claiming a systematic variation of several galaxy properties with offset from the main-sequence; the off-sequence galaxies show more compact structures (\citealt{Elbaz+11}; \citealt{Wuyts+11b}), warmer interstellar dust (\citealt{Magnelli+14}), higher gas-to-M$_*$ ratio (\citealt{Gao&Solomon04}), larger Far-to-Mid Infrared flux ratio (\citealt{Nordon+10}, \citeyear{Nordon+12}) and higher star formation efficiency (SFE\footnote{SFE is defined as the ratio between SFR and cold gas mass.}, \citealt{Daddi+10b}; \citealt{Genzel+10}). The question of whether the transition from MS to starburst galaxies is steady or discontinuous remains open.

A similar two fold-scheme is found also for the star-forming properties of low-luminosity X-ray-selected AGN (L$_{\rm X}<$10$^{44}$ erg s$^{-1}$), showing at best a weak correlation between L$_{\rm X}$ and SFR, while bright Quasars follow a positive correlation with SFR at least up to z$\sim$1, probably driven by major mergers (\citealt{Lutz+08}; \citealt{Netzer09}; \citealt{Shao+10}; \citealt{Lutz+10}; \citealt{Rosario+12}). At z$\sim$2 such a correlation seems weak or absent (\citealt{Rosario+12}; \citealt{Harrison+12}). However, compelling evidences of AGN-driven feedback (e.g. \citealt{Farrah+12}) and the inverted correlation found in smaller samples of luminous z$\sim$2 AGN (\citealt{Page+12}) caution that our current picture of AGN/galaxy coevolution is still dependent on sample statistics and selection biases.

By exploiting large samples of X-ray selected AGN, \citet{Mullaney+12a} found that about 80 per cent of X-ray AGN live in main-sequence galaxies, 5--10 per cent in starburst galaxies, and about 10--15 per cent in quiescent systems. While \citet{Santini+12} reported larger mean SFRs for the hosts of X-ray AGN compared to a mass-matched inactive reference that includes both star-forming and passive systems, \citet{Rosario+13} found very similar SFRs in X-ray AGN hosts and a mass-matched reference of only star-forming galaxies. All three studies reinforce the idea that most of the SMBH accretion is taking place in star-forming systems. 

In contrast to the studies of AGN hosts and to reach a comprehensive understanding of the cosmic SMBH growth, several recent studies take a census of AGN accretion history on the basis of Far Infrared (FIR) and/or mass-selected samples of galaxies. Unlike the weak or absent correlation between star formation and black hole accretion rate (BHAR\footnote{The terms BHAR and $\langle \dot{M}_{\rm bh} \rangle $ adopted throughout the paper are assumed to have the same physical meaning.}) for samples of AGN hosts, there is clear correlation of average BHAR and key properties of galaxy samples. Positive and close to linear correlation is found between average BHAR and mean stellar mass at various redshifts (\citealt{Mullaney+12b}) as well as with SFR (\citealt{Rafferty+11}; \citealt{Chen+13}). These apparently contradictory results have been interpreted by \citet{Hickox+14} as due to different variability time scales between nuclear activity and global star formation. According to this scenario, both components are 
intimately connected at any time: while star formation is relatively stable over $\sim$100 Myr, the AGN might vary over $\sim$5 orders of magnitude on very short (about 10$^5$ yr) time scales (\citealt{Hickox+09}; \citealt{Aird+12}; \citealt{Bongiorno+12}). In this scenario, all episodes of star formation are accompanied by SMBH growth, but only when smoothing over the variations of individual sources do the average properties of AGN and their hosts show a consistent evolution, as stated by \citet{Mullaney+12b} and \citet{Chen+13}.

Both these latter studies derived average trends by binning their parent samples as a function of M$_*$ or SFR. Since their selection techniques were mostly sensitive to main-sequence galaxies, in principle the resulting correlations found with average BHAR might be primarily due to one parameter, but reflected into a correlation with the other one, simply because of the main-sequence relation that holds between the two. To break this degeneracy and investigate in detail the role of AGN accretion in the context of galaxy evolution, it is necessary to split the sample as a function of both SFR and M$_*$ and study the evolution of the average AGN accretion properties in the SFR--M$_{*}$ plane at different redshifts.

The primary goal of this work is to map the average BHAR as a function of SFR, M$_{*}$ and redshift. Our analysis exploits one of the widest compilations of FIR selected galaxies at non-local redshifts. Robust SFRs for each individual source of the sample are measured from data taken by the \textit{Herschel Space Observatory} (\citealt{Pilbratt+10}). Our sample spans about three orders of magnitude in M$_*$ and four in SFR in the redshift range 0$<$z$\leq$2.5. For the first time we also investigate the role of AGN activity in off-sequence galaxies with respect to their main-sequence counterparts, seeking to constrain the parameter that primarily drives the growth of active SMBHs.

We used the FIR data in COSMOS (\citealt{Scoville+07}) and from the Great Observatories Origins Deep Survey (GOODS) South (GOODS-S) and North (GOODS-N) fields, obtained with the \textit{Herschel}-Photodetector Array Camera and Spectrometer (PACS, \citealt{Poglitsch+10}), as part of the PACS Evolutionary Probe (PEP\footnote{\url{http://www2011.mpe.mpg.de/ir/Research/PEP/} }, \citealt{Lutz+11}) project. In the GOODS fields, PEP data are also combined with the deepest observations of the GOODS-\textit{Herschel} (GOODS-H\footnote{\url{http://hedam.oamp.fr/GOODS-Herschel} }; \citealt{Elbaz+11}) open time key program. In addition, PACS observations at 70 (in GOODS-S only), 100 and 160$~\mu$m are supplemented with sub-millimeter photometry at 250, 350 and 500$~\mu$m obtained by the Spectral and Photometric Imaging Receiver (SPIRE, \citealt{Griffin+10}), as part of the \textit{Herschel} Multi-tiered Extragalactic Survey (HerMES\footnote{\url{http://hermes.sussex.ac.uk}}, \citealt{Oliver+12}).

The paper is structured as follows. In Section \ref{sample} we present our parent sample and multi-wavelength photometry. In Section \ref{sfr_mass_plane} we introduce individual \textit{Herschel} sources in the SFR--M$_*$ plane, while their average X-ray properties are derived in Section \ref{xray-analysis}. We present the observed relationships between AGN and galaxy properties in Section \ref{results}, and discuss the implication of this work in Section \ref{discussion}. We list our concluding remarks in Section 7. Throughout this paper, we assume a \citet{Chabrier03} initial mass function (IMF) and a flat cosmology with $\Omega_m = 0.30$, $\Omega_\Lambda = 0.70$, and $H_0 = 70$ km\,s$^{-1}$\,Mpc$^{-1}$.

\section{Sample Description} \label{sample}

Our sample exploits \textit{Herschel}-PACS (Data Release 1) and SPIRE observations in the GOODS-North (GN hereafter), GOODS-South (GS hereafter) and COSMOS fields, covering in total about 2 deg$^2$. In the following Sections we present the parent sample and briefly mention the cross-match with multi-wavelength identifications, referring the reader to \citet{Lutz+11}; \citet{Berta+11}; \citet{Oliver+12} and \citet{Magnelli+13} for a detailed description of data reduction and construction of multi-wavelength catalogues.

\subsection{Far-Infrared selected galaxies} \label{fir-sample}

The parent sample includes all sources in the GOODS and COSMOS fields with $>$3$\sigma$ flux density in at least one PACS band. 
In both GOODS fields, FIR data are taken from the blind catalogues described in \citet{Magnelli+13} that combine the data of PEP (\citealt{Lutz+11}) and GOODS-H (\citealt{Elbaz+11}). In the GOODS fields, flux density limits (3$\sigma$) reach $\approx$ 1--2 mJy in the PACS bands, depending on filter and depth of observation, and $\approx$ 8 mJy in SPIRE-250$~\mu$m. The confusion limit reachable with PACS ranges between 1.3 and 5 mJy (5$\sigma$, \citealt{Berta+11}), while SPIRE-250$~\mu$m observations are fully limited by confusion noise (\citealt{Oliver+12}). In the COSMOS field, the depth achieved by SPIRE-250$~\mu$m observations is more comparable to that reached by PACS ones, being $\approx$ 5, 8 and 10 mJy at 100, 160 and 250$~\mu$m, respectively. This is the main reason why we only used PACS-selected galaxies in the GOODS fields, while in COSMOS our selection also exploits SPIRE-selected ones. We note that the background level of PACS and SPIRE observations is relatively flat in each field. We checked 
that galaxies taken from the SPIRE-selected catalogue have consistent SFR and M$_*$ values (within a factor of two) with those taken from the PACS-selected sample at the same redshift. The presence of SPIRE-selected sources allows to double our sample of star-forming galaxies without introducing a significant bias in our analysis. The extraction of PACS flux densities in all fields was performed blindly as described in \citet{Berta+11}, \citet{Lutz+11} and \citet{Magnelli+13}, while for SPIRE sources it follows the approach presented by \citet{Roseboom+11}.

\subsection{Multi-wavelength identification} \label{multi-lambda}

The cross-match between PEP data and the extensive broad-band photometry available from the ultraviolet (UV) to the sub-millimeter has been accomplished via a maximum likelihood algorithm (\citealt{Sutherland+92}; \citealt{Ciliegi+01}) to deep Multiband-Imaging Photometer (MIPS) \textit{Spitzer} detections at 24$~\mu$m (\citealt{Magnelli+11}, \citeyear{Magnelli+13}), whose positions have been used as priors to extract SPIRE fluxes in the sub-mm (\citealt{Roseboom+11}). The cross-match to optical/UV wavelengths in both GOODS fields is described in detail by \citeauthor{Berta+10} (\citeyear{Berta+10}, \citeyear{Berta+11}). We collected 892 (GS) and 850 (GN) FIR-selected sources with broad-band photometry from the optical/UV to the sub-mm.

An extensive photometric coverage is available also in COSMOS, where fluxes from the PEP catalogue have been cross-matched to 24$~\mu$m data (\citealt{LeFloc'h+09}), in turn used as positional priors to get SPIRE fluxes (\citealt{Roseboom+11}) and then matched to the optically-based catalogues from \citet{Capak+07} and \citet{Ilbert+09}. The same algorithm has been adopted for SPIRE-250$~\mu$m sources with no PACS detection. Totally, the number of FIR sources in COSMOS with either PACS or SPIRE-250$~\mu$m detection is about 17000.

Given that optical and near-IR observations in these fields are deeper than Far-IR ones obtained from \textit{Herschel}, the fraction of \textit{Herschel}-selected galaxies without a counterpart in optical/near-infrared catalogues reaches only a few per cent in each field.

\subsubsection{X-ray counterparts} \label{xray-catalogues}
 
We used optical/near-infrared counterpart positions to cross-match our \textit{Herschel}-selected sample with available X-ray data from \textit{Chandra} observations in COSMOS and in the GOODS fields.

\begin{itemize}
 \item In the GOODS-South, we use the 4-Ms \textit{Chandra}-Deep Field South (CDF-S) observations (\citealt{Xue+11}).  The X-ray catalogue provides count rates and observed fluxes for each source in different bands: soft (0.5--2 keV), hard (2--8 keV) and full (0.5--8 keV). 

 \item In the GOODS-North, X-ray data are available from the 2-Ms \textit{Chandra}-Deep Field North (CDF-N) observations (\citealt{Alexander+03}). 

 \item In the COSMOS field, observations from the \textit{Chandra}-COSMOS (C-COSMOS, \citealt{Elvis+09}; \citealt{Civano+12}) survey are publicly available but only cover about 0.9 deg$^2$ (reaching about 160 ks in the central 0.45 deg$^2$ and 80 ks outside) instead of $\sim$2 deg$^2$ scanned by \textit{Herschel}. Consequently, we limit the parent sample to match the common sky area, which implies a cut to $\sim$45 per cent of the original population, leading from roughly 17000 to 7272 FIR-selected galaxies. 

\end{itemize}

After having collected as many spectroscopic redshifts as possible (see Section \ref{redshift}), we cut our original sample at redshift z$\leq$2.5, since above this threshold poor statistics would affect the significance of our results. This leads to 829, 804 and 7011 sources in GS, GN and COSMOS, respectively, for a total number of 8644 FIR-selected galaxies across all fields. More than 90 per cent of these FIR-selected galaxies are detected in at least two \textit{Herschel} bands. The cross-match with X-ray detections has been made via a neighborhood algorithm, by assuming 1 arcsec matching radius between optical positions of the counterparts assigned to X-ray and \textit{Herschel} sources, respectively. The number of X-ray detections is 212/829 in GS, 134/804 in GN and 448/7011 in COSMOS.

As highlighted in previous studies (e.g. \citealt{Rosario+12}), we confirm that the fraction of \textit{Herschel} sources detected in X-rays is generally small and is a function of field, depending on the relative depth of X-ray/IR observations. 
The fraction of \textit{Herschel} sources detected by \textit{Chandra} is 26 per cent in the GOODS-S, 17 per cent in the GOODS-N and only $\sim$6 per cent in COSMOS. In all fields, we have taken the observed (i.e. obscured) X-ray fluxes in each band (soft, hard and full) from the publicly available catalogues.

We derived the average X-ray properties for the rest of the sample by performing a stacking analysis on X-ray maps (see Section \ref{stacking} and Appendix \ref{appendix_A}).

\subsection{Spectroscopic and photometric redshifts } \label{redshift}

\begin{figure}
\begin{center}
    \includegraphics[width=80mm]{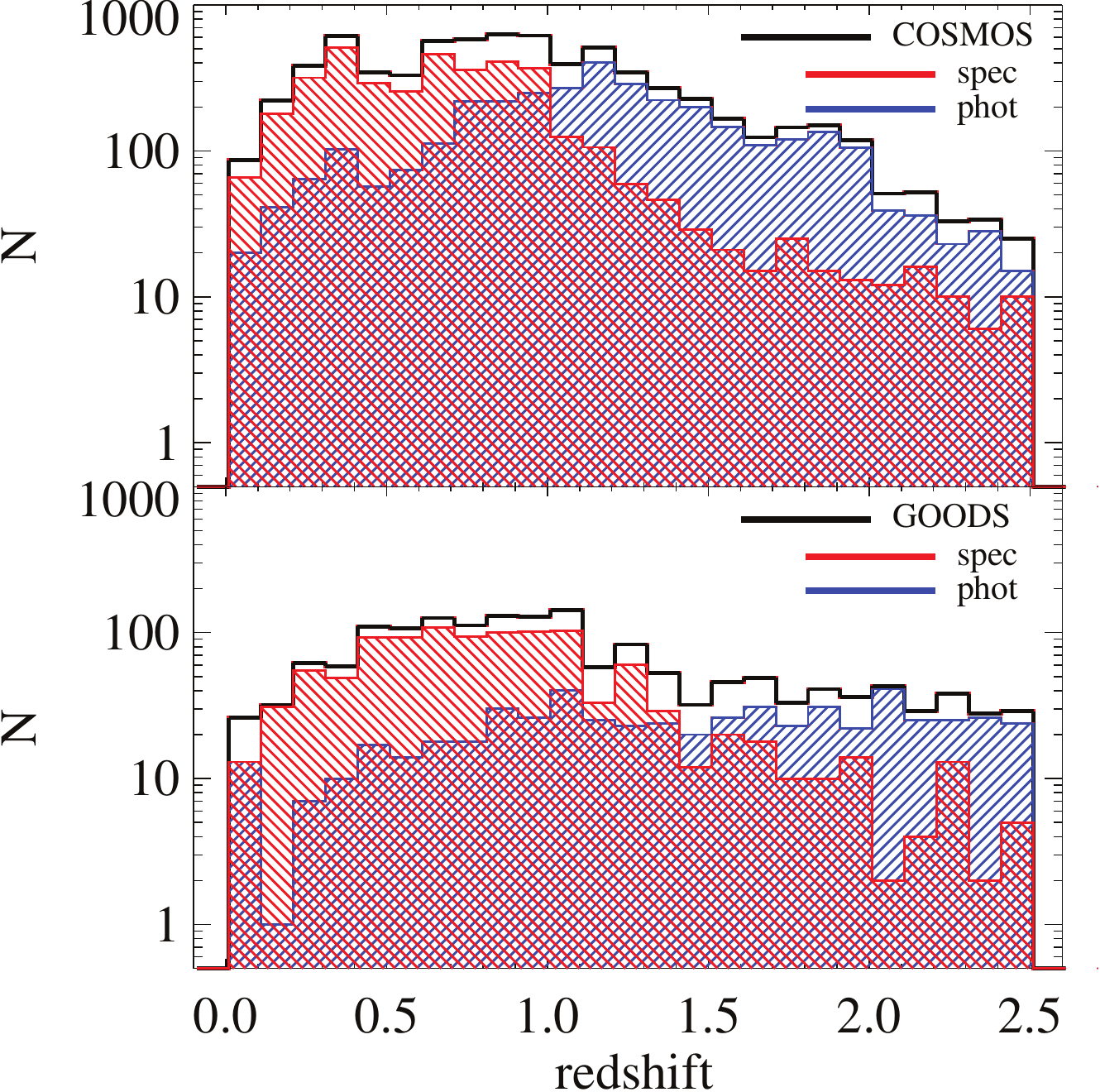}
\end{center}
 \caption{Redshift distribution of \textit{Herschel} galaxies in COSMOS (top panel) and GOODS (bottom panel). Spectroscopic and photometric redshifts are shown in red and blue, respectively, while the black line is the sum of the two. Note that the scale of the \textit{y}-axis is logarithmic.}
   \label{fig:z_dist}
\end{figure}

Extensive redshift compilations are publicly available in both GOODS and COSMOS. In Fig. \ref{fig:z_dist} the redshift distributions for the GOODS and COSMOS samples are shown, distinguishing between spectroscopic (red) and photometric (blue) measurements. We defer the reader to \citet{Berta+11} for a careful description of redshift catalogues, including uncertainties on photometric redshifts. Here we just provide a short list of references of redshift measurements. The redshift completeness for our \textit{Herschel} sources with counterpart in optical/near-infrared catalogs reaches 100 per cent in all fields.

\begin{itemize}

  \item GOODS: in GOODS-S, we extended the original spectroscopic sample presented by \citet{Grazian+06} and \citet{Santini+09} in the GOODS MUlti-wavelength Southern Infrared Catalog (GOODS-MUSIC) with publicly available spectroscopic redshifts, as described in \citet{Berta+11}, reaching a global spectroscopic fraction for our \textit{Herschel} sample as high as 67 per cent. In GOODS-N, redshift measurements are taken from \citet{Berta+11}, who collected spectroscopic redshifts from \citet{Barger+08} for about 64 per cent of the \textit{Herschel}-selected sample. In both fields, photometric redshifts have been derived by \citet{Berta+11} by using the EAZY (\citealt{Brammer+08}) code, as described in \citet{Wuyts+11a}. 
  
  \item COSMOS: we used photometric redshifts from \citet{Ilbert+10} and \citet{Wuyts+11a}. For \textit{Chandra} detected sources we have made a cross-check with photometric redshifts presented by \citet{Salvato+11} which are more suitable for AGN dominated sources. We retrieved spectroscopic measurements from the $z$COSMOS survey by \citeauthor{Lilly+07} (\citeyear{Lilly+07}, \citeyear{Lilly+09}), either the public $z$COSMOS-bright or the proprietary $z$COSMOS-deep data base. We also browsed the most recent public spectroscopic surveys, replacing our photometric redshifts with spectroscopic ones in case of high reliability: \citet{Ahn+14}, from Data Release 10 of the Sloan Digital Sky Survey (SDSS); \citet{Coil+11} from the PRIsm MUlti-object Survey (PRIMUS) catalogue; \citet{Trump+09} from the COSMOS-Magellan spectroscopic catalogue. Globally, about 50 per cent of the COSMOS FIR-selected sample have a spectroscopic redshift.

\end{itemize}

The overall fraction of spectroscopic redshifts is larger than 60 per cent at z$\leq$1.5, while it decreases to 20 per cent at z$\sim$2. We stress that, if limiting our sample to spectroscopic redshifts only, all the results would remain consistent within 1$\sigma$ uncertainty with those already presented and discussed in Sections \ref{results} and \ref{discussion}.

\begin{figure*}
\begin{center}
    \includegraphics[width=140mm,keepaspectratio]{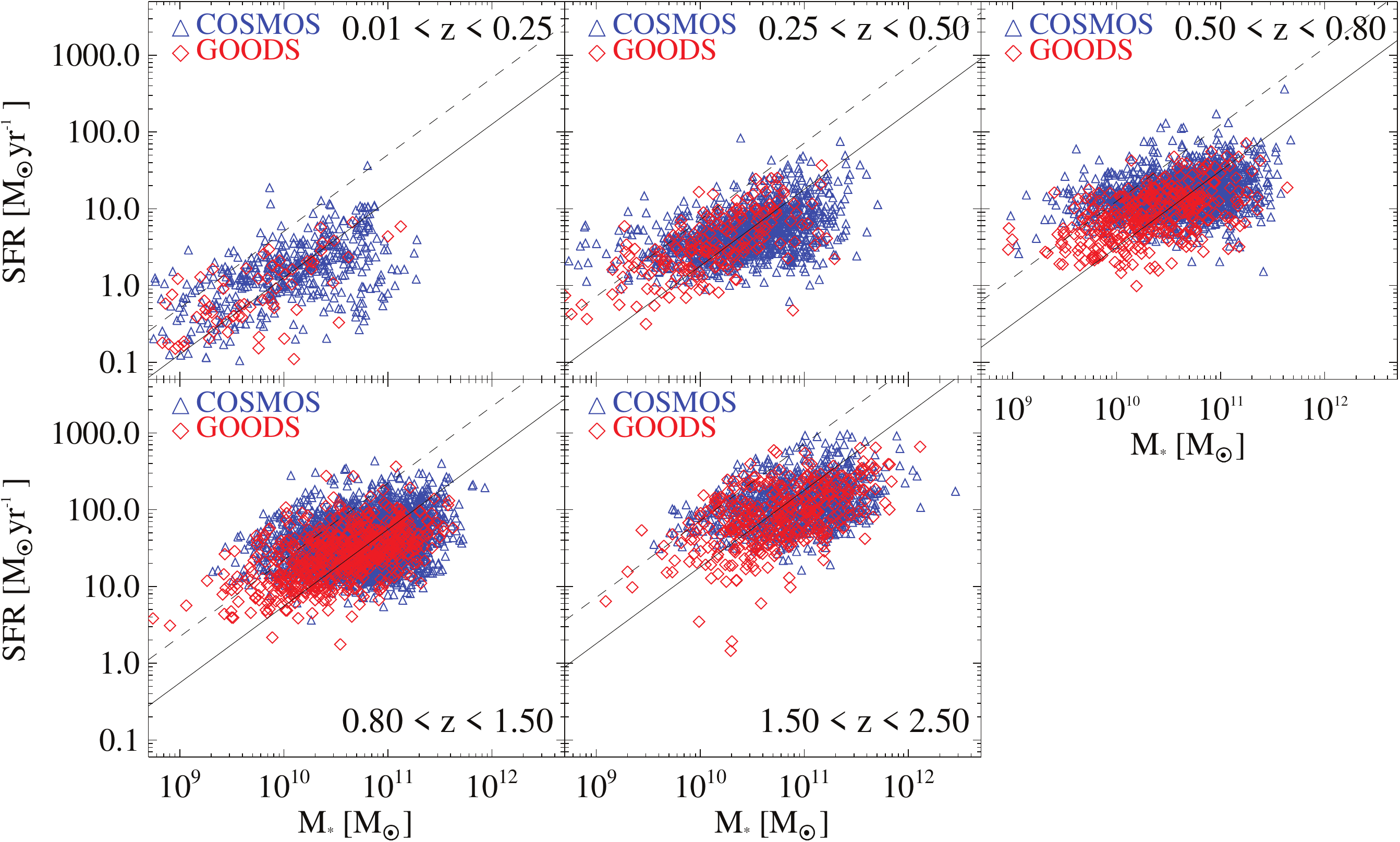}
\end{center}
 \caption{Individual \textit{Herschel}-selected sources as a function of SFR, M$_{*}$ and redshift. Red and blue symbols refer to GOODS-S/N and COSMOS sources, respectively. The black solid line at any redshift represents the main-sequence defined by \citet{Elbaz+11}. The black dashed line marks 4 $\times$ higher sSFR as a threshold between main and off-sequence galaxies.}
   \label{fig:sfr_mass_plane}
\end{figure*}

\section{The SFR--M$_{*}$ plane} \label{sfr_mass_plane}

Given the wealth of photometric datapoints available in all fields from the UV to the sub-mm, we performed broad-band SED decomposition to derive M$_{*}$ and SFR for the entire sample. Each observed SED has been fitted with the {\sc MAGPHYS}\footnote{{\sc magphys} can be retrieved at \url{http://www.iap.fr/magphys/magphys/MAGPHYS.html} } code (\citealt{daCunha+08}), as well as with a modified version of {\sc MAGPHYS} adapted to include an AGN component (\citealt{Berta+13}) using AGN templates by \citet{Fritz+06} and \citet{Feltre+12}. The best-fit obtained with the AGN is preferred if the resulting $\chi^2$ value significantly decreases (at $\geq$99 per cent confidence level, on the basis of a Fisher test) compared to the fit without the AGN. In this case, SFRs and M$_{*}$ estimates are taken from the fit with AGN, otherwise from the original {\sc magphys} code. However, even if the AGN component is required in the best-fit, we stress that its contribution to the galaxy IR luminosity is not dominant (i.e. $\
lesssim$10 per cent) for most of the sample. We defer the reader to \citet{Berta+13} for further details on SED decomposition, and to \citet{Delvecchio+14} for statistical analysis.

The SFR has been derived by converting the total IR (rest 8-1000$~\mu $m) luminosity taken from the best-fit galaxy SED (i.e. corrected for a possible AGN emission) using the conversion from \citet{Kennicutt98}, scaled to a \citet{Chabrier03} IMF\footnote{We computed the SFR for each galaxy by accounting for its obscured star formation only. As pointed out by \citet{Magnelli+13}, the fraction of unobscured SFR density ranges between 12 per cent and 25 per cent at z$<$2, but drops to a few per cent for FIR-selected samples of galaxies (e.g. \citealt{Wuyts+11a,Wuyts+11b}).}. The M$_*$ is derived from the SED decomposition itself, which allows to get robust measurements also for type-1 AGN, as most of them show near-IR ($\approx$1$~\mu$m) emission dominated by the host galaxy light (e.g. \citealt{Bongiorno+12}). We have checked that our estimates of M$_*$ for optically identified type-1 AGN are consistent within the uncertainties (around 0.3 dex) with those presented by \citet{Bongiorno+12}, with no systematics.

The sample has been split in five different redshift bins: 0.01$\leq$z$<$0.25, 0.25$\leq$z$<$0.50, 0.50$\leq$z$<$0.80, 0.80$\leq$z$<$1.50 and 1.50$\leq$z$\leq$2.50. We place our galaxies on the SFR--M$_{*}$ plane in Fig. \ref{fig:sfr_mass_plane}, marking with different colours GOODS (red) and COSMOS (blue) sources. Given that PACS flux density limits in the GOODS fields are about 5 times lower than in COSMOS, \textit{Herschel} observations in the GOODS fields detect fainter IR galaxies (i.e. lower SFR) compared to observations in COSMOS at the same redshift. Since our \textit{Herschel}-based selection is sensitive to the most star-forming galaxies in each field, the wedge traced by the observed galaxy distribution in the SFR--M$_{*}$ is relatively flat (see \citealt{Rodighiero+11}) compared to the linear main-sequence relation defined by \citet{Elbaz+11}. The MS evolution with redshift is parametrized as sSFR$_{\rm ms}$ = 26 $\times$ t$_{\rm cosmic}^{-2.2} \rm [Gyr^{-1}]$ (\citealt{Elbaz+11}), where t$_{\rm 
cosmic}$ is the cosmic time (in Gyr) starting from the Big Bang. Our selection also includes off-sequence galaxies, with sSFR $>4\times$sSFR$_{\rm ms}$ ($\sim$0.6 dex), corresponding to symbols above the black dashed lines. In each redshift slice, the sample has been split in both SFR and M$_{*}$, taking 0.5$\times$0.5 dex bins\footnote{We note that the typical uncertainty on individual SFR and M$_*$ measurements is of the order of 0.2--0.3 dex, so significantly smaller than the bin-width.}. Instead of keeping a fixed binning configuration in SFR and M$_{*}$ at all redshifts, we decided to center our bins on the main-sequence relation in each bin of M$_{*}$ and redshift\footnote{We checked also for alternative binning configurations (i.e. either different shapes for individual bins, or different placement on the SFR--M$_{*}$ plane) but this did not produce any significant impact on our results.}. This arrangement is preferable, since it allows us to better highlight any potential systematics in 
terms of average AGN properties between main and off-sequence galaxies.

\section{X-ray analysis} \label{xray-analysis}

In this Section we present the X-ray analysis of our \textit{Herschel} selected sample. For sources detected in X-ray ($\sim$ 10 per cent) we get count rates, fluxes and observed luminosities in different bands from the above-mentioned catalogues (section \ref{xray-catalogues}). X-ray undetected sources represent most ($\sim$ 90 per cent) of the \textit{Herschel} sample studied in this work. To derive their average X-ray properties, we stack the X-ray maps (section \ref{stacking}), in the observed soft, hard and full bands. Subsequently, we characterize the average X-ray properties of \textit{Herschel} galaxies in each bin of SFR, M$_{*}$ and redshift by considering X-ray detections and stacks together (section \ref{joined}). Then we subtracted the X-ray flux expected from star formation (section \ref{subtraction}) in single X-ray bands and corrected the remaining X-ray emission for the nuclear obscuration (section \ref{obscuration}) to derive the intrinsic (unobscured) mean AGN X-ray luminosity (rest-frame).
 
Using widely-adopted conversion factors, we employ the final nuclear X-ray luminosity as a proxy to evaluate the mean BHAR in each bin of SFR, M$_*$ and redshift (see Section \ref{results}).

\subsection{Stacking the X-ray maps} \label{stacking}

Here we briefly mention the main steps concerning the X-ray stacking and refer the reader to Appendix \ref{appendix_A} for further details.
We combined all \textit{Herschel} sources undetected in X-rays from both GOODS and COSMOS fields and we grouped them as a function of SFR, M$_{*}$ and redshift. After masking all X-ray detected sources which could potentially affect the stacked signals, we piled up single cutouts of X-ray undetected sources in the same bin of SFR, M$_{*}$ and redshift, centered on their optical coordinates. For each object, we defined regions from which we extract source and background photons, in order to derive background-subtracted (i.e. net) photon counts. Finally, we needed to correct for differential sensitivities between various X-ray fields, so we normalized the resulting net photon counts of each object by the corresponding effective (i.e. corrected for instrumental effects) exposure time, which provides exposure-corrected mean count rates in each observed X-ray band\footnote{We note that our results are in good agreement, within the uncertainties, with those obtained using CSTACK (\url{http://cstack.ucsd.
edu/cstack/}, developed by T. Miyaji) on the same X-ray maps.}. To convert count rates into observed fluxes, we assumed a power-law spectrum with $\Gamma$=1.4, based on the empirically observed spectrum of the X-ray background (e.g. \citealt{Gilli+07}).

\subsection{Mean X-ray luminosity of \textit{Herschel} sources} \label{joined}

To get the overall X-ray properties of our \textit{Herschel}-selected galaxies, we collected in each bin of SFR, M$_*$ and redshift both X-ray detected and undetected sources. Assuming that the bin includes $N$ sources, $m$ detected and $n$ undetected in X-rays, we computed a number weighted average of their X-ray fluxes, according to the following formula:

\begin{equation}
 \left \langle  S_{\rm bin} \right \rangle = \frac{n \times S_{\rm stack} + \sum_{i=1}^{m} S_i }{N}  
     \label{eq:average}
\end{equation}
where $S_{\rm stack}$ is the mean X-ray flux obtained by stacking $n$ undetected sources and S$_i$ is the individual flux measured for the $i$-th detected source within the same bin. We stress that the above-mentioned expression is a number weighted average, therefore appropriate to investigate \textit{mean} properties (e.g. mean X-ray luminosity) of the underlying galaxy population, but we caution that it does not necessarily represent the most \textit{probable} value of the observed distribution of a given parameter. Considering that $\approx$90 per cent of \textit{Herschel} galaxies are not X-ray detected, as well as the fact that X-ray detections are about 50 times brighter than stacks, in most bins X-ray detected and stacked sources provide comparable contributions to the mean X-ray flux. Rest-frame average X-ray luminosities in each band, for both X-ray detected and stacked sources, are derived from mean X-ray fluxes by assuming a power-law spectrum with intrinsic slope $\Gamma$=1.9 (\citealt{Tozzi+06};
 \citealt{Mainieri+07}) and no obscuration. 
We performed this calculation as a function of SFR, M$_{*}$ and redshift.

\subsection{Subtraction of X-ray emission from star formation} \label{subtraction}

Using X-rays to investigate the level of AGN activity in the SFR--M$_*$ plane requires the subtraction of X-rays from other processes in the host galaxy, especially those related to star formation. This subtraction will allow us to apply the correction for nuclear obscuration (Section \ref{obscuration}) only to the AGN-related X-ray emission rather than the total (i.e. AGN + galaxy) one. It is known that X-ray observations provide a relatively clean selection of accreting SMBHs, but a fraction of the X-ray emission (e.g. in \textit{Chandra} Deep Fields, CDFs) may also come from X-ray binaries and the hot interstellar medium (e.g. \citealt{Mineo+12a,Mineo+12b}). A single threshold (i.e. $\rm L_X$ = 3$\times$10$^{42}$ erg s$^{-1}$) in X-ray luminosity is not a proper way of classifying these galaxy populations, since it is known that X-ray emission from star-forming galaxies shows a positive correlation with SFR. Previous works (e.g. \citealt{Ranalli+03}) calibrated this relation in the local Universe, while 
recent studies (\citealt{Vattakunnel+12}; \citealt{Mineo+14}; \citealt{Symeonidis+14}) uncovered this relation up to z$\sim$1.5. 

Though all these relations are calibrated through independent analyses and selection techniques, they provide reasonably consistent (within a factor of 2) estimates of the X-ray emission expected from star formation. However, we prefer the relation by \citet{Symeonidis+14}, since they also exploited \textit{Herschel} data and performed stacking on X-ray maps to better characterize the average L$\rm _X$--SFR correlation in inactive (i.e. non-AGN) SFR-selected galaxies. They found a quasi linear relation holding at 1$<$SFR$<$1000 M$_{\odot}$ yr$^{-1}$ and not evolving significantly with redshift up to z$\sim$1.5. We scaled their relation to a \citet{Chabrier03} IMF and converted to our rest-frame soft (0.5--2 keV) and hard (2--8 keV) X-ray bands:

\begin{equation}
 \rm L_X^{\rm soft} \rm[erg ~ s^{-1}] = 2.04 \times 10^{39} ~ \rm SFR ~ \rm [\rm M_{\odot}~yr^{-1}]
   \label{symeonidis1}
\end{equation}
\begin{equation}
 \rm L_X^{\rm hard} \rm[erg ~ s^{-1}] = 5.13 \times 10^{39} ~ \rm SFR ~ \rm [\rm M_{\odot}~yr^{-1}]
   \label{symeonidis2}
\end{equation}
In each bin we subtracted the X-ray luminosity due to star formation from the average X-ray luminosity (section \ref{joined}), both in the soft and the hard band. However, we note that this subtraction does not impact significantly the previous values, as the typical non-AGN contribution arising from the joint (X-ray detected + stacked) sample is less than 10 per cent.

\subsection{Correction for nuclear obscuration} \label{obscuration}

To derive the intrinsic AGN luminosity in each bin of SFR, M$_*$ and redshift, we considered the AGN-related X-ray emission derived in Section \ref{subtraction} and followed the approach developed by \citet{Xue+10}, who used the hardness ratio (HR) as a proxy to estimate the nuclear obscuration. The hardness ratio is defined as:
\begin{equation}
\rm HR = \rm (CR_{hard} - CR_{soft}) / (CR_{hard} + CR_{soft}) 
\end{equation}
where $\rm CR_{hard}$ and $\rm CR_{soft}$ represent the (exposure-corrected) count rates in hard and soft-bands, respectively. 

Given that we are not able to constrain the HR for each individual galaxy of the sample, we correct the mean AGN X-ray emission of all \textit{Herschel} galaxies (both detected and undetected in X-rays, see Eq. \ref{eq:average}) for an average level of obscuration. We make the simple assumption that the mean HR calculated after the subtraction in soft and hard X-ray bands (Eqs. \ref{symeonidis1}, \ref{symeonidis2}) is representative of the galaxy population in the same bin of SFR, M$_*$ and redshift.

We parametrized the effect of nuclear obscuration by assuming a single power-law X-ray spectrum, with intrinsic photon index $\Gamma=1.9$ (model wa$\times$zwa$\times$po in XSPEC, \citealt{Arnaud96}) and accounting for absorption, both Galactic and intrinsic to the AGN. However, since the hardness ratio deals with photon counts instead of fluxes, one needs to convolve the intrinsic model with the instrument response curve.\footnote{The instrument response curve has been corrected for several instrumental effects: vignetting, exposure time variations, energy-dependent efficiency, variation of effective area with time of observations.} After performing this convolution, we built a set of simulated spectra that predict the observed hardness ratio as a function of hydrogen column density ($\rm N_H$) and redshift. We calculated the observed hardness ratio from AGN count rates obtained in Section \ref{subtraction} and selected the spectral model (i.e. intrinsic column density) best reproducing the observed HR at 
the mean redshift of the underlying galaxy population in the same bin of SFR and M$_*$. From the observed-frame, absorption-corrected fluxes in the 0.5--8 keV band ($S_{\rm [0.5-8],int}$), we calculated the rest-frame, intrinsic full-band X-ray luminosity as follows:

\begin{equation}
 \rm L_{\rm [0.5-8],int} = 4\pi ~ D_L^2 ~ S_{\rm [0.5-8],int} ~ (1+z)^{\Gamma-2}
    \label{eq:lx_corr}
\end{equation}
where $\rm D_L$ is the luminosity distance corresponding to the mean redshift of the \textit{Herschel} population in a given bin, while the intrinsic photon index $\Gamma$ is set to 1.9.

As a sanity check, we also compared our obscuration-corrected full-band (0.5--8 keV) luminosities with those presented by \citet{Xue+11} in the \textit{Chandra}-Deep field South (CDF-S) 4-Ms catalogue and found an excellent agreement, as expected, given that we followed similar approaches. We found that obscuration level does not significantly affect the average AGN X-ray luminosity, as the typical correction factor is about 1.3. The obscuration-corrected X-ray luminosities estimated through hardness ratio are generally consistent within a factor of $\sim$30 per cent with more precise measurements from spectral--fitting analysis (\citealt{Xue+11}). Nevertheless, as argued by \citet{Xue+11}, the level of obscuration might be affected by strong uncertainties in case of highly obscured AGN. In addition, we compared our predictions based on HR with spectral measurements of $\sim$400 \textit{Chandra}-COSMOS AGN presented by \citet{Lanzuisi+13}. We ended up with reasonably small scatter (about 0.2 dex) in [0.5--8] 
keV X-ray luminosities, even for sources classified as highly oscured ($\rm N_H > 10^{23}~$cm$^{-2}$) AGN from X-ray spectral--fitting. Given these sanity checks and the relatively small average corrections for obscuration obtained for our sample, this might suggest that the average intrinsic X-ray luminosity of \textit{Herschel}-selected galaxies does not arise primarily from highly obscured AGN. However, a thorough X-ray spectral analysis of these sources would be beyond the scope of this paper.

\subsubsection{Uncertainty on the [0.5--8] keV intrinsic L$\rm _X$} \label{errors}

We evaluated the uncertainty on the [0.5--8] keV intrinsic X-ray luminosity $\rm L_{\rm [0.5-8],int}$ by performing a bootstrapping analysis. This technique provides reliable error bars, especially in case a small fraction of the objects dominate the signal. 

Suppose there are $N$ objects populating a given bin of SFR, M$_*$ and redshift, out of which $m$ are detected in X-rays, while $n$ are not detected. We selected at random $N$ objects from the same bin, allowing duplication of the same source. This random extraction likely leads to different numbers of detections ($m'$) and non-detections ($n'$). We combined the photon counts together from the new $n'$ sources to derive stacked count rates and fluxes. Then we applied Eq. \ref{eq:average} (see Section \ref{joined}) to get the average X-ray flux representative of this random realization. With 1000 iterations of this calculation, we obtained the distribution of average count rates and fluxes of the galaxy sample. The 16$^{th}$ and 84$^{th}$ percentiles of the final distribution set the 1$\sigma$ lower and upper bounds on the measured X-ray flux. 

To evaluate the uncertainty on the obscuration-corrected L$\rm_X$, we iterated 1000 times the same analysis described in Sections \ref{subtraction} and \ref{obscuration}, once for each random realization. This approach returns the final distribution of the [0.5--8] keV intrinsic X-ray luminosities, with $\pm$1$\sigma$ error bars estimated through bootstrapping. Since this technique would provide quite large error bars (i.e. almost unconstrained fluxes and luminosities) in case of poor statistics, we have required a minimum number of sources in each bin ($\geq 15$, regardless of the number of detections and non-detections).

\section{Results} \label{results}

\begin{figure*}
\begin{center}
    \includegraphics[width=140mm,keepaspectratio]{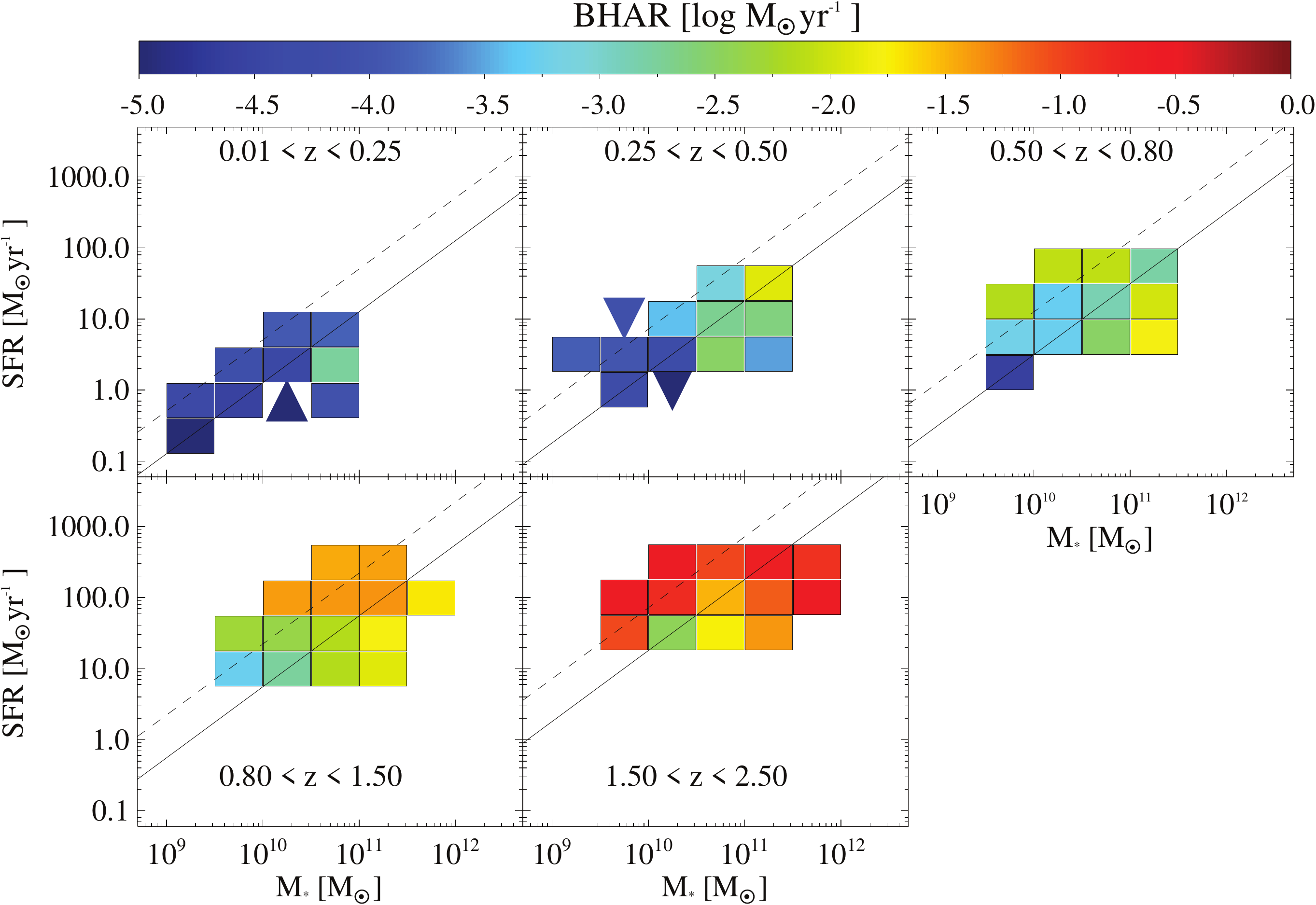}
\end{center}
 \caption{BHAR distribution (colour-coded) in the SFR--M$_*$ plane at 0.01$\leq$z$\leq$2.5. The bins in SFR and M$_*$ are arranged to sample the main-sequence relation (black solid line), which evolves with redshift. The black dashed line divides main-sequence from off-sequence galaxies. Coloured bins include at least 15 sources, either detected or undetected in X-rays. Upward and downward triangles set 1$\sigma$ lower and upper limits, respectively, on the average BHAR. In particular, as the expected non-AGN contribution was subtracted from the average X-ray luminosities obtained in Section \ref{joined}, in some bins this subtraction left solely the hard (soft) X-ray emission, which returned a lower (upper) limit in the final intrinsic X-ray luminosity. }
   \label{fig:plane_agn}
\end{figure*}

We used the intrinsic X-ray luminosity as a proxy for mapping the average BHAR in the SFR--M$_*$ plane and studying its correlation with integrated galaxy properties.

\subsection{Average BHAR} \label{average_bhar}

Obscuration-corrected X-ray luminosities in each bin have been turned into average bolometric AGN luminosities $\langle \rm L_{\rm bol} \rangle $ by assuming a set of luminosity-dependent X-ray bolometric corrections from \citet{Marconi+04}\footnote{We remark that if taking a fixed bolometric correction value of 22.4 (as done by \citealt{Mullaney+12b} and \citealt{Chen+13}), the AGN bolometric luminosities would be larger by a factor of about 2, for L$_{\rm X}<$10$^{44}$ erg s$^{-1}$.}. We have assumed a constant conversion factor to convert the average AGN bolometric luminosity $\langle \rm L_{\rm bol} \rangle $ (in erg s$^{-1}$) to average SMBH accretion rate $ \langle \dot{M}_{\rm bh} \rm(M_{*},SFR,z) \rangle $ (in M$_{\odot}$yr$^{-1}$), according to the formula (see \citealt{Alexander+12}):

\begin{equation}
 \langle \dot{M}_{\rm bh} \rm(M_{*},SFR,z) \rangle = 0.15 ~ \frac{\epsilon}{0.1} ~ \frac{ \langle L_{\rm bol} (\rm M_{*}, \rm SFR, z) \rangle }{10^{45} }
\end{equation}
where the matter-to-radiation conversion efficiency $\epsilon$ is assumed to be 10 per cent (e.g. \citealt{Marconi+04}). Table \ref{tab:data} lists redshift, SFR, M$_*$ and accretion rates for all bins involved, as well as the number of sources entering each bin. In Fig. \ref{fig:plane_agn} we show the SFR--M$_*$ plane at different redshifts with colour-coded average BHAR. A few upward and downward triangles set 1$\sigma$ lower and upper limits, respectively. They replace the formal values in case the correction for obscuration is not applicable. Indeed, the subtraction of the star-formation X-ray emission from the total (i.e. AGN and star-formation) X-ray luminosity obtained in Section \ref{joined}, in some bins left solely the hard (soft) X-ray emission, which returned a lower (upper) limit in the [0.5--8] keV intrinsic X-ray luminosity.

In addition, an upper limit is also imposed in case the observed X-ray spectrum is softer than any spectral model with intrinsic $\Gamma$=1.9\footnote{The softest X-ray spectrum that is reproducible with intrinsic slope $\Gamma$=1.9 corresponds to HR = --0.47. }. Indeed, this slope represents just a mean value among the overall distribution of AGN X-ray spectra, whose typical scatter is around 0.2 (e.g. \citealt{Piconcelli+05}). We note that our SFR--M$_*$--z grid follows the evolution with redshift of the MS. Even at our large source statistics, uncertainties in BHAR remain noticeable (about 0.4 dex on average, see Table \ref{tab:data}), also due to the fact that in some bins a few, intrinsically luminous X-ray detections might dominate the global signal, as stated in Equation \ref{eq:average}. 

In Fig. \ref{fig:plane_agn}, the high SFR Herschel galaxies at higher redshift show larger average BHAR than the more local and lower SFR galaxies. Also, within the panels for each redshift slice, trends are indicated with SFR and/or M$_*$. We proceed to study which of these galaxy properties correlates best with BHAR, and discuss results in the context of 0$<$z$\leq$2.5 galaxy evolution.

\begin{table*}
   \caption{Best-fit parameters returned by fitting a linear relation in the $\log$--$\log$ space, as a function of redshift, between $\langle \dot{M}_{\rm bh} \rangle $ and galaxy properties ($x$-parameter). $\alpha$ and $\beta$ are the slope and intercept of the linear best-fit (see Eq. \ref{eq:linfit}). Values in brackets set the 1$\sigma$ uncertainty on the related parameters. The Spearman's rank correlation coefficient $\rho$ is the strength of the correlation, while P(correlation) represents the significance of its deviation from zero, considering the number of points $N$ in each panel. In columns 5 and 6 their estimates have been derived from a linear regression fit, while in columns 7 and 8 we used a partial-correlation fitting (see text for details). }
\begin{tabular}{lcccccc cc}
\hline
(1) $z$-bin & (2) $x$--parameter & (3) $\alpha$ & (4) $\beta$ & (5) $\rho$ & (6) P(correlation) & (7) partial $\rho$  & (8) partial P(correlation) & (9) $N$ \\
\hline
0.01 $\leq$ z $<$ 0.25 & SFR [M$_{\odot}~$yr$^{-1}$] & 0.54 ($\pm$0.27) & --4.31 ($\pm$0.12) & 0.68 & 95.8\% & 0.39  &  83.0\%  & 9  \\
0.25  $\leq$ z $<$ 0.50 & SFR [M$_{\odot}~$yr$^{-1}$] & 1.06 ($\pm$0.24) & --4.16 ($\pm$0.38) & 0.75 & 99.1\% &  0.49  &   92.6\% &  11  \\
0.50  $\leq$ z $<$ 0.80 & SFR [M$_{\odot}~$yr$^{-1}$] & 1.05 ($\pm$0.52) & --4.02 ($\pm$0.64) & 0.60 & 96.1\% &  0.40  &   89.9\% &  12  \\
0.80  $\leq$ z $<$ 1.50 & SFR [M$_{\odot}~$yr$^{-1}$] & 1.44 ($\pm$0.30) & --4.29 ($\pm$0.46) & 0.80 & 99.9\% &  0.71  &   99.6\% &  14  \\
1.50  $\leq$ z $\leq$ 2.50 & SFR [M$_{\odot}~$yr$^{-1}$] & 1.13 ($\pm$0.38) & --3.42 ($\pm$0.78) & 0.68 & 99.0\% &  0.68  &   99.2\% & 13  \\
\hline
0.01  $\leq$ z $<$ 0.25 & M$_*$ [M$_{\odot}$] & 0.44 ($\pm$0.21) & --8.63 ($\pm$2.17) & 0.73 & 97.5\%  &  0.52  &   91.7\% & 9  \\
0.25  $\leq$ z $<$ 0.50 & M$_*$ [M$_{\odot}$] & 0.76 ($\pm$0.33) & --11.41 ($\pm$3.51) & 0.75 & 99.1\%  &  0.52  &   93.8\% & 11  \\
0.50  $\leq$ z $<$ 0.80 & M$_*$ [M$_{\odot}$] & 1.11 ($\pm$0.34) & --14.59 ($\pm$3.66) & 0.58 & 95.2\%  &   0.41 &   89.9\% & 12  \\
0.80  $\leq$ z $<$ 1.50 & M$_*$ [M$_{\odot}$] & 1.07 ($\pm$0.18) & --13.58 ($\pm$1.96) & 0.65 & 98.8\%  &   0.43 &   93.0\% & 14  \\
1.50  $\leq$ z $\leq$ 2.50 & M$_*$ [M$_{\odot}$] & 0.25 ($\pm$0.27) & --3.89 ($\pm$2.94) & 0.23 & 56.3\% &   --0.17 &   30.0\% & 13  \\
\hline
0.01  $\leq$ z $<$ 0.25 & sSFR/sSFR$_{\rm ms}$ & --0.09 ($\pm$0.27) & --4.17 ($\pm$0.13) & --0.02 & 33.2\%  &  /   & /   & 9  \\
0.25  $\leq$ z $<$ 0.50 & sSFR/sSFR$_{\rm ms}$ & --0.18 ($\pm$0.42) & --3.33 ($\pm$0.21) & --0.17 & 43.6\%  & /   &  /  & 11  \\
0.50  $\leq$ z $<$ 0.80 & sSFR/sSFR$_{\rm ms}$ & --0.53 ($\pm$0.50) & --2.74 ($\pm$0.17) & 0.03 & 36.7\%   &  /  &  /  & 12  \\
0.80  $\leq$ z $<$ 1.50 & sSFR/sSFR$_{\rm ms}$ & --0.53 ($\pm$0.40) & --2.13 ($\pm$0.15) & 0.19 & 54.6\%   & /   &  /  & 14  \\
1.50  $\leq$ z $\leq$ 2.50 & sSFR/sSFR$_{\rm ms}$ & 0.10 ($\pm$0.31) & --1.10 ($\pm$0.15) & 0.40 & 83.8\%  &  / & /  & 13  \\
\hline

\end{tabular}

\label{table:fit}
\end{table*}

\subsection{Correlation of BHAR with galaxy properties} \label{correlations}

We explore here the observed trends between $\langle \dot{M}_{\rm bh} \rangle $ and various galaxy properties: SFR, M$_*$ and offset from the MS\footnote{As the adopted MS relation shows a linear trend at all redshifts, the offset from the MS becomes simply the ratio between the sSFR and that corresponding to the main-sequence sSFR$_{\rm ms}$, for a given bin of M$_*$ and redshift.}. Because of the evolution with cosmic time of SFRs and BHARs, we fit the data separately for each redshift bin, by assuming a linear trend in the $\log$--$\log$ space through:
\begin{equation}
\log (\langle \dot{M}_{\rm bh} \rangle ) = \alpha \times \log (x) + \beta
    \label{eq:linfit}
\end{equation}
where $\alpha$ and $\beta$ represent slope and intercept, respectively, while $x$ is the corresponding independent variable. We have used the IDL routine {\sc imsl\_multiregress.pro}, which performs a linear regression fit considering error bars in both variables, and returns best-fit intercept and slope with related 1$\sigma$ uncertainties. The linear best-fits obtained in each redshift bin are summarized in Table \ref{table:fit}.

\subsubsection{Spearman's rank} \label{spearman}

To evaluate the significance of the observed trends, we used the Spearman's rank correlation coefficient $\rho$, that indicates the strength of any observed correlation. However, the Spearman's test considers the observed data points as ``exact'' and does not take into account their possible error bars. To estimate the most probable correlation coefficient given the error bars, we used the procedure detailed in \citet{Santos-Sanz+12} (see their Appendix B.2 for details). Briefly, they generated 1000 samples of data points, building each synthetic dataset from its associated Gaussian distribution function, where error bars correspond to one standard deviation. This Monte Carlo bootstrap analysis provides a distribution of Spearman's rank correlation coefficients, where its median value gives the weighted Spearman's coefficient that represents the most likely correlation to the observed data points. We used the IDL routine {\sc r\_correlate.pro} to derive the average value of Spearman's $\rho$: in addition, 
this function provides the significance of its deviation from zero $\rm P(correlation)$ of the observed trend, considering the number of points populating the correlation. These parameters are used to compare different relationships and infer the strength of the observed correlations. We applied this analysis when studying the relationship between average BHAR and SFR, M$_*$ and MS--offset in each redshift bin (see Table \ref{table:fit}).

\begin{figure}
\begin{center}
    \includegraphics[width=\linewidth]{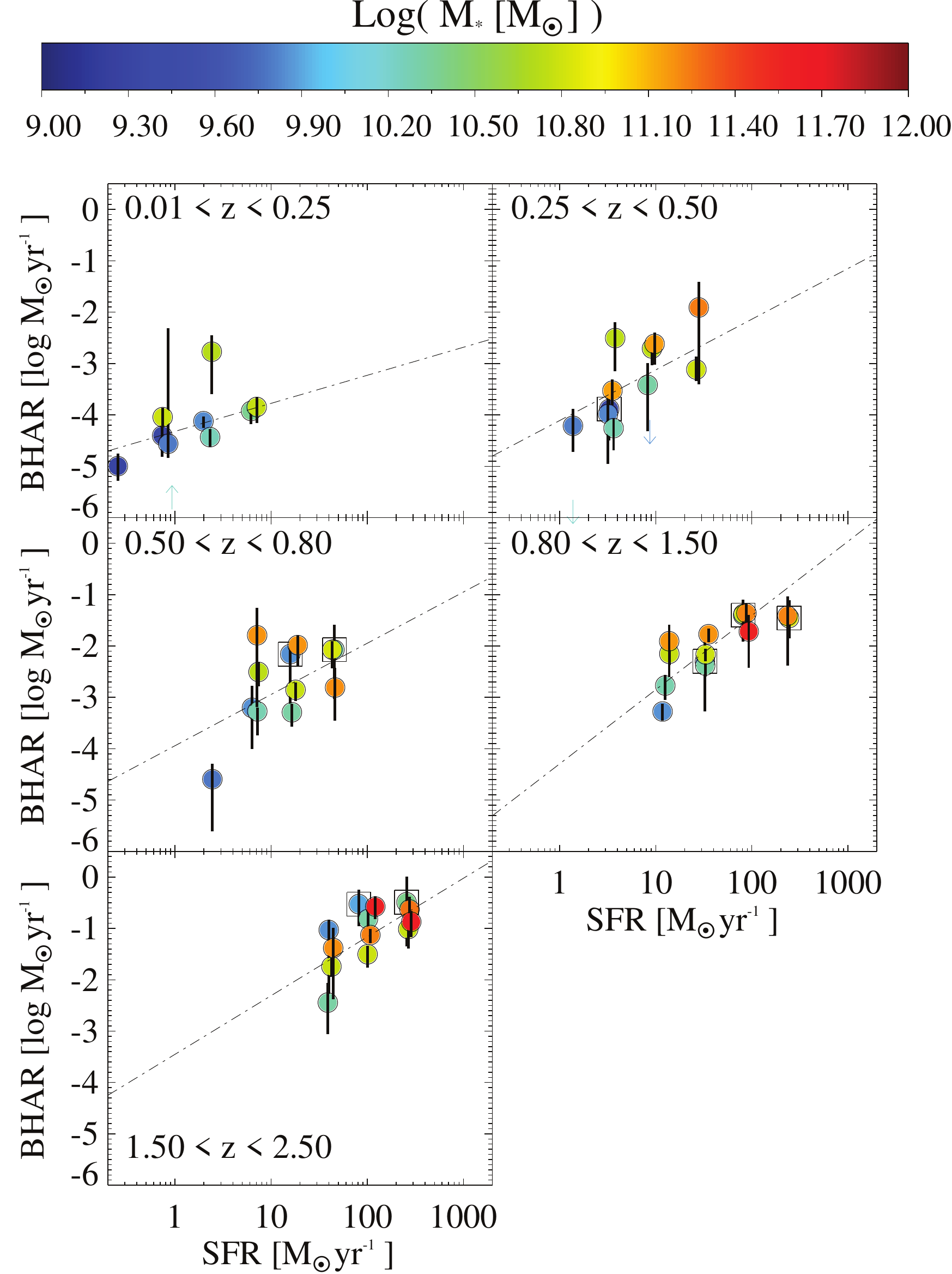}
\end{center}
 \caption{Average BHAR vs SFR, at different redshifts. The colour-coded bar is the M$_*$, while individual black dashed-dotted lines show the best-fit linear relation inferred in each $z$-bin, whose coefficients are listed in Table \ref{table:fit}. Bins representing off-sequence galaxies are enclosed in black squares. We stress that the lowest redshift bin might suffer from incompleteness at high SFRs.}
   \label{fig:agn_sfr}
\end{figure}

\begin{figure}
\begin{center}
    \includegraphics[width=\linewidth]{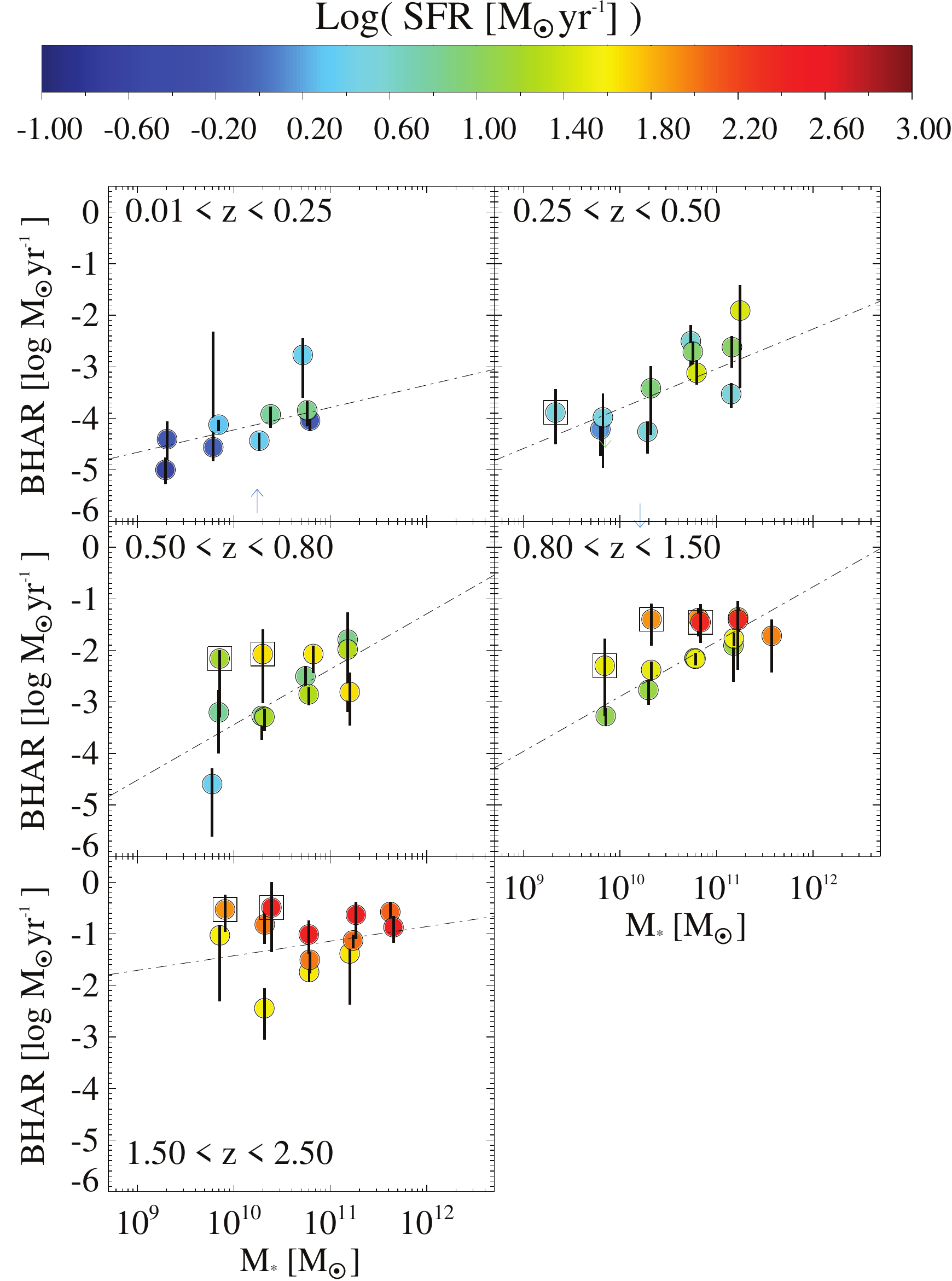}
\end{center}
 \caption{Average BHAR vs M$_*$, at different redshifts. The colour-coded bar is the SFR, while individual black dashed-dotted lines set the best-fit linear relation inferred in each $z$-bin, whose coefficients are listed in Table \ref{table:fit}. Bins representing off-sequence galaxies are enclosed in black squares. We stress that the lowest redshift bin might suffer from incompleteness at high M$_*$.}
   \label{fig:agn_m*}
\end{figure}

\subsubsection{Correlation of BHAR with SFR and M$_*$} \label{corr_sfr_m*}

Fig. \ref{fig:agn_sfr} and \ref{fig:agn_m*} show the relations $\langle \dot{M}_{\rm bh} \rangle $ vs SFR and $\langle \dot{M}_{\rm bh} \rangle $ vs M$_*$, respectively, in different redshift bins. Circles correspond to the same bins already shown in Fig. \ref{fig:plane_agn}, but projected on the $\langle \dot{M}_{\rm bh} \rangle $--SFR or the $\langle \dot{M}_{\rm bh} \rangle $--M$_*$ plane. Error bars of each bin mark the $\pm$1$\sigma$ uncertainties in $\langle \dot{M}_{\rm bh} \rangle $, obtained by applying a Monte Carlo bootstrapping (see Section \ref{errors}) to the [0.5--8] keV intrinsic X-ray luminosity and rescaling to $\langle \dot{M}_{\rm bh} \rangle $ as described in Section \ref{average_bhar}. The black dashed dotted line in each $z$-bin represents the best-fit linear relation parametrized in Eq. \ref{eq:linfit}, whose parameters are listed in Table \ref{table:fit}. The best-fit slope ($\alpha$) suggests that the correlation between BHAR and SFR is consistent with a linear relation, at z$\geq$0.
25. 
Fig. \ref{fig:agn_sfr} and \ref{fig:agn_m*} show that the best-fit relations at 0.01$\leq$z$<$0.25 are flatter than linear. This is likely due to the fact that the comoving volume covered by our fields is not large enough to detect Ultra Luminous InfraRed Galaxies (ULIRGs, L$_{\rm IR} > $10$^{12} \rm L_{\odot}$) hosting powerful Quasars. At z$>$0.25 both relationships show clear trends with nearly linear slopes and relatively high significance ($>$2$\sigma$ level). Focusing on individual redshift slices, the comparison between Spearman's rank coefficients in Table \ref{table:fit} (columns 5 and 6) shows that the strength of the observed correlations of BHAR with either SFR or M$_*$ are comparable at z$<$0.8, while at higher redshifts the correlation with SFR becomes progressively more significant. Especially at z$\sim$2, these coefficients significantly favour the correlation with SFR, and show a nearly flat trend between $\langle \dot{M}_{\rm bh} \rangle $ and M$_*$. 

Another way of illustrating this point is by looking at the z$>$0.5 panels in Fig. \ref{fig:agn_sfr} and \ref{fig:agn_m*}: Fig. \ref{fig:agn_m*} shows that off-sequence galaxies (highlighted with black squares) have higher average BHAR, compared to mass-matched main-sequence galaxies. In contrast, at a given SFR in Fig. \ref{fig:agn_sfr}, the BHAR for off-sequence galaxies does not stand out of those for main-sequence galaxies. Overall, this may indicate that at z$>$0.5 the BHAR is primarily dependent on SFR. 

Given the well known main-sequence relation between SFR and M$_*$ at all redshifts considered here, it is necessary to carry this analysis further and simultaneously investigate the dependence of $\langle \dot{M}_{\rm bh} \rangle $ on both SFR and M$_*$. We addressed this issue by performing a partial-correlation fitting analysis. We used the IDL routine {\sc r\_correlate.pro} to evaluate the Spearman's $\rho$ related to each couple of parameters and then we combined them according to the following expression:
\begin{equation}
 \rho_{\rm ab\dot c} = \rm \frac{\rho_{ab} - \rho_{ac} \rho_{bc}}{\sqrt{ (1-\rho_{ac}^2) (1-\rho_{bc}^2) }} 
\end{equation}
which returns the partial correlation between $a$ and $b$ adjusted for $c$. Assuming that (\textit{a,b,c}) = (SFR, $\langle \dot{M}_{\rm bh} \rangle $, M$_*$), then $\rm \rho_{ab\dot c}$ would represent the partial correlation between SFR and $\langle \dot{M}_{\rm bh} \rangle $, removing the concomitant dependence on M$_*$. As done in Section \ref{spearman}, we performed again a Monte Carlo bootstrap analysis to provide a distribution of Spearman’s coefficients from partial-correlation analysis, taking the median value of that distribution as the most probable to reproduce our data\footnote{If taking the mean value instead of the median would change the resulting $\rho$ by around 2--3 per cent. Moreover, we note that, if the observed trend is poorly significant, the most probable $\rho$ value might prefer a mildly negative rather than a mildly positive correlation (see Table \ref{table:fit}), or viceversa.}. The latter $\rho$ values obtained from partial-correlation fitting (herafter ``partial $\rho$''), 
together with their corresponding significance levels, are listed in Table \ref{table:fit} (columns 7 and 8).

\begin{table}
   \caption{Best-fit parameters obtained in each $z$-bin through multiple-correlation fitting procedure. $\gamma$ is the slope of SFR, while $\delta$ refers to M$_*$. Numbers in brackets represent their 1$\sigma$ uncertainties. The last column provides with the reduced $\chi^2$ values returned from the multiple linear fitting.}
\centering 
\begin{tabular}{lcccc}

\hline
$ ~~~~~ z$-bin & $\gamma$ (SFR) & $\delta$ (M$_*$) &  $\chi^2_{\nu}$ \\
\hline
0.01  $\leq$ z $<$ 0.25 & 0.38 ($\pm$0.20) & 0.31 ($\pm$0.16) & 1.90   \\
0.25  $\leq$ z $<$ 0.50 & 0.78 ($\pm$0.32) & 0.52 ($\pm$0.24) & 1.78   \\
0.50  $\leq$ z $<$ 0.80 & 0.48 ($\pm$0.36) & 0.92 ($\pm$0.29) & 1.93   \\
0.80  $\leq$ z $<$ 1.50 & 0.85 ($\pm$0.22) & 0.73 ($\pm$0.16) & 0.89   \\
1.50  $\leq$ z $\leq$ 2.50 & 1.16 ($\pm$0.33) & --0.04 ($\pm$0.18) &  1.86  \\
\hline
\end{tabular}

\label{table:fit2}
\end{table}

\begin{figure}
\begin{center}
    \includegraphics[width=\linewidth]{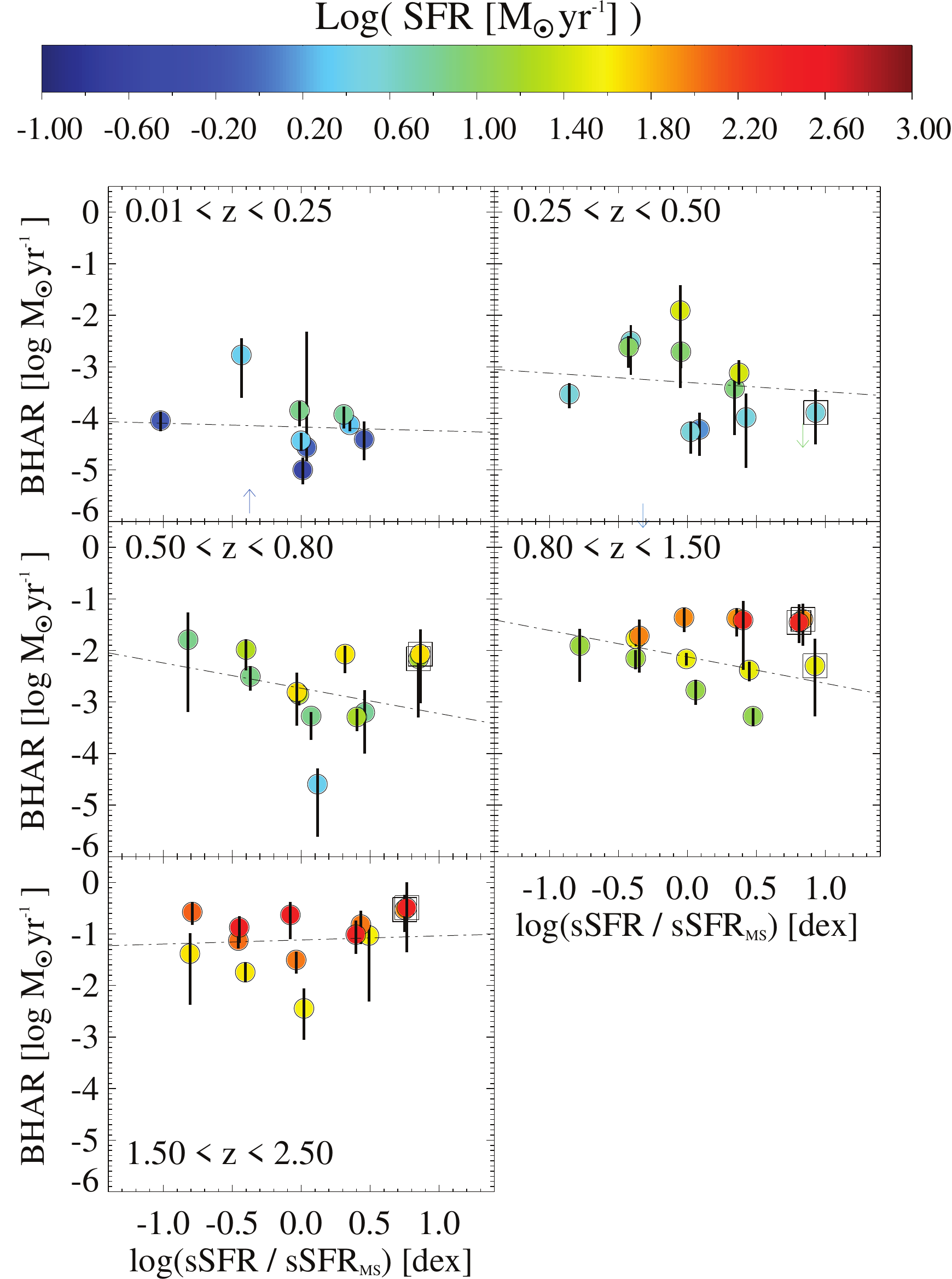}
\end{center}
 \caption{Average BHAR vs offset from the MS, at different redshifts. The colour-coded bar is the SFR, while individual black dashed-dotted lines set the best-fit linear relation inferred in each $z$-bin, whose coefficients are listed in Table \ref{table:fit}. Bins representing off-sequence galaxies are enclosed in black squares.}
   \label{fig:agn_offset}
\end{figure}

The comparison between partial $\rho$ values and those obtained from a rank correlation of individual parameters shows similar trends. This suggests that the main-sequence relation between SFR and M$_*$ does not significantly affect the correlations with the average BHAR. The observed trends show that a correlation with M$_*$ is slightly preferable than with SFR at z$<$0.5, while at higher redshifts the correlation with SFR becomes progressively more significant. However, we evaluated the difference between SFR and M$_*$, in terms of resulting partial-correlation coefficients, to be less than 1$\sigma$ at z$<$0.8 and of the order of 2.2$\sigma$ at 1.5$<$z$<$2.5. This suggests that for our current sampling of the SFR--M$_*$ plane, the overall evolution of the average BHAR is best represented through a joint dependence on both SFR and M$_*$ at all redshifts. A difference is detected only in the highest \textit{z}-bin, where the trend with SFR is fairly preferable with respect to that with M$_*$. 

We performed a multiple-correlation fitting in the $\log$ space to provide a simple analytic expression of the simultaneous dependence of $\langle \dot{M}_{\rm bh} \rangle $ on both SFR and M$_*$ as a function of redshift. According to the following parametrization:
\begin{equation}
 \log (\langle \dot{M}_{\rm bh} \rangle) \propto \rm \gamma \log(SFR) +  \delta \log(M_*)
     \label{eq:partial}
\end{equation}
we defined $\gamma$ and $\delta$ as the slopes of SFR and M$_*$, respectively, and listed their values in Table \ref{table:fit2} for each $z$-bin. At z$<$0.8, the slopes related to SFR and M$_*$ are always consistent with each other (within 1$\sigma$ uncertainty), meaning that both parameters are required to trace the evolution of the average BHAR. As expected from our previous analyses, at z$\sim$2 the correlation with $\langle \dot{M}_{\rm bh} \rangle $ is primarily driven by the SFR.

Given that we are studying \textit{Herschel}-selected galaxies, our sample is biased towards the most star-forming galaxies at any given redshift. Since this selection effect could potentially bias our results, we tried to evaluate the implications that incompleteness effects might have on this study. We repeated the same analysis by limiting our parent sample to \textit{Herschel} galaxies with FIR flux (in either PACS or SPIRE bands) larger than the flux corresponding to 80 per cent completeness level. This cut strongly reduces our statistics (about a factor of three), as well as the range of galaxy properties we are allowed to probe, but it allows us to test the validity of our findings in a reasonably complete sample of star-forming galaxies. We repeated the same analysis and evaluated the strength of the observed correlations between average BHAR, SFR and M$_*$, on the basis of this quite complete sample of star-forming galaxies. We ended up with consistent trends (within 1$\sigma$ uncertainty) in all 
redshift bins, with overall correlations between $\langle \dot{M}_{\rm bh} \rangle $ and galaxy properties in agreement with our previous analysis. In addition, we checked that the observed trends obtained separately in GOODS fields and in COSMOS provide consistent results. 

We have found in this Section that average BHAR correlates with SFR and with M$_*$ for star-forming galaxies in a wide range of SFRs and redshifts. SFR may be the primary driver of these trends, at least for z$>$0.8. A further investigation will be feasible through forthcoming \textit{Chandra} observations of the entire COSMOS field from the COSMOS Legacy Survey (PI: F. Civano), by which we plan to increase statistics and significance of the correlations in future work.

\subsubsection{Correlation of BHAR with offset from the MS} \label{corr_offset}

Given long-standing evidence about AGN in local massive above-MS objects (ULIRGs), it is natural to probe for a link of BHAR and offset from the MS. This is shown in Fig. \ref{fig:agn_offset} and the best-fit parameters obtained through a linear regression fitting are listed in Table \ref{table:fit}. It is evident that the average BHAR does not depend significantly on the offset from the MS. At all redshifts, the best-fit slope is consistent with a flat or slightly declining relation. Our analysis based on Spearman's $\rho$ coefficients suggests a very weak, or absent, correlation at any redshift here considered\footnote{We have verified that our results remain if using the MS definition of \citet{Whitaker+12} or \citet{Schreiber+14}, instead of the one proposed by \citet{Elbaz+11}.}. 

If we focus our analysis only to the few bins corresponding to above-MS objects (black squares in Fig. \ref{fig:agn_m*}), they tend to show, at a given M$_*$, larger BHAR compared to galaxies placed on the MS relation at the same redshift. However, this increase of BHAR from main-sequence to off-sequence galaxies is not observed when considering all bins, probably due to the wider range in specific SFR covered by MS galaxies in our sample. In addition, we have looked into possible trends of the ratio BHAR/SFR with offset from the MS and found no significant one at any redshift. 

In Section \ref{discussion} we discuss this lack of trends in the framework of classical merger scenarios. All these findings are consistent with a link of BHAR to SFR, irrespective of whether a given SFR is reached in a bursting above-MS galaxy of moderate M$_*$, or a high M$_*$ main sequence galaxy.

\subsection{BHAR as a function of SFR and M$_*$ over 0.01$\leq$z$\leq$2.5} \label{bhar_all}

\begin{figure}
\begin{center}
$\begin{array}{cc}
         \includegraphics[width=\linewidth]{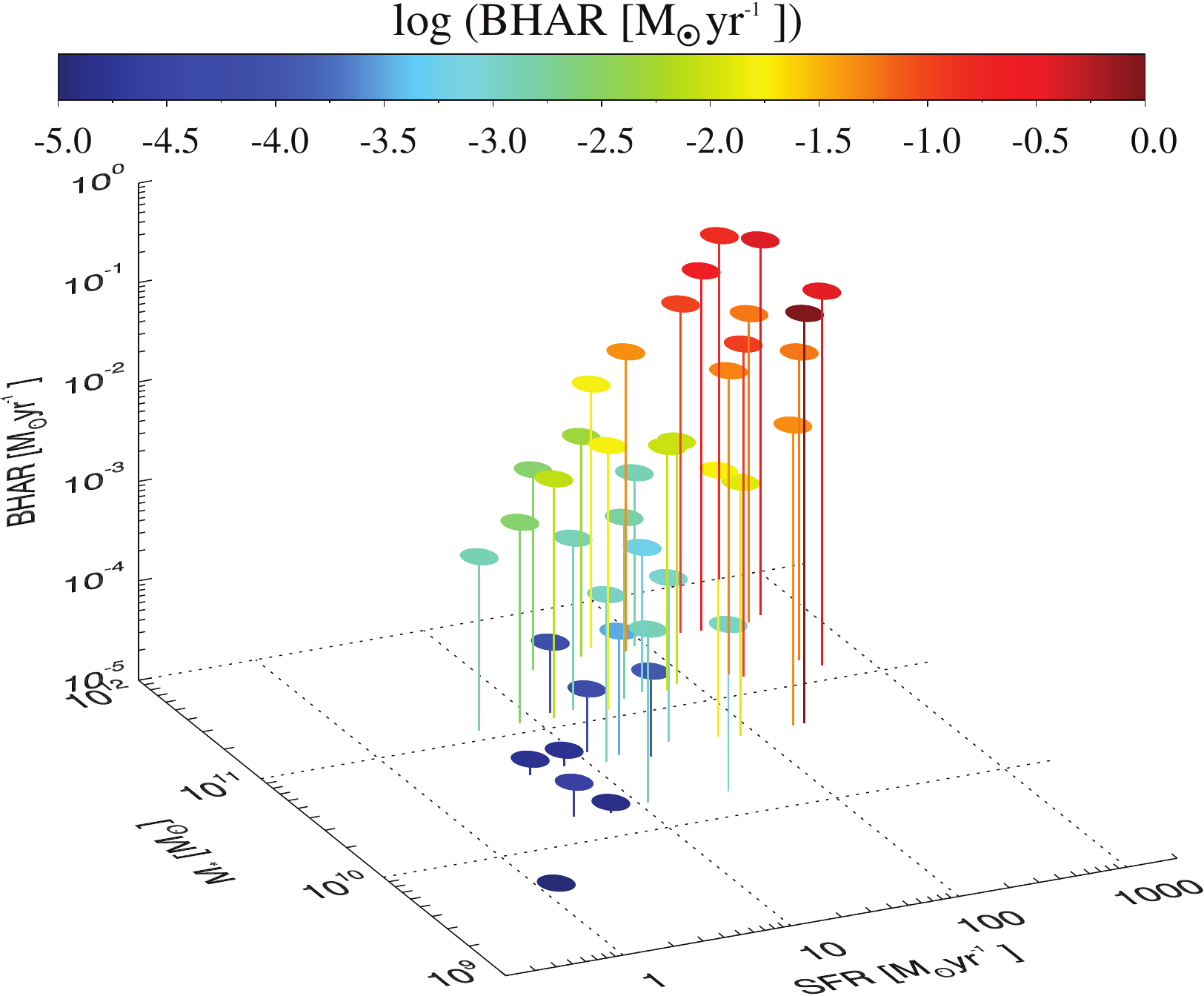} \\
         \includegraphics[width=3.2in]{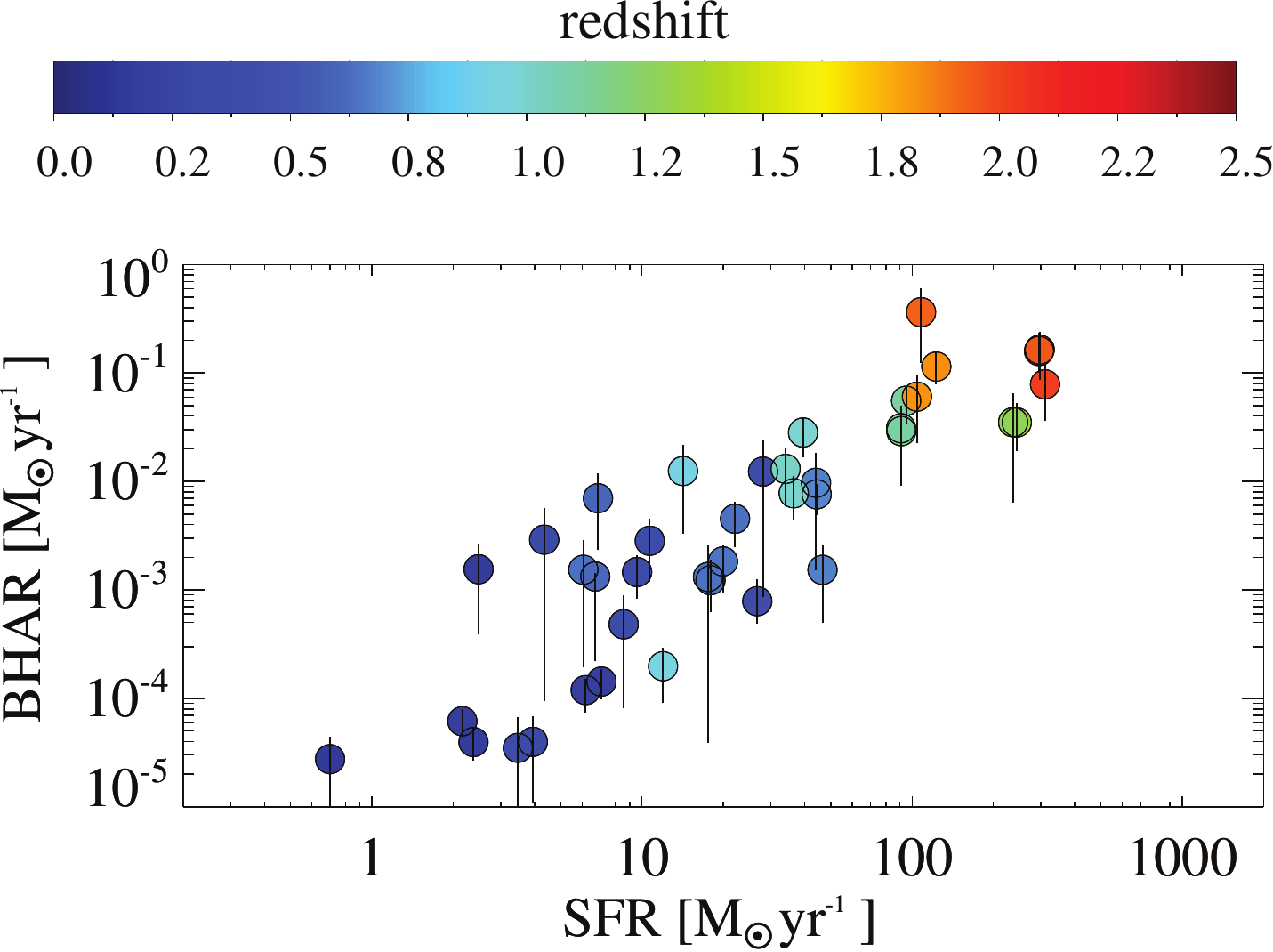} \\
         \includegraphics[width=3.2in]{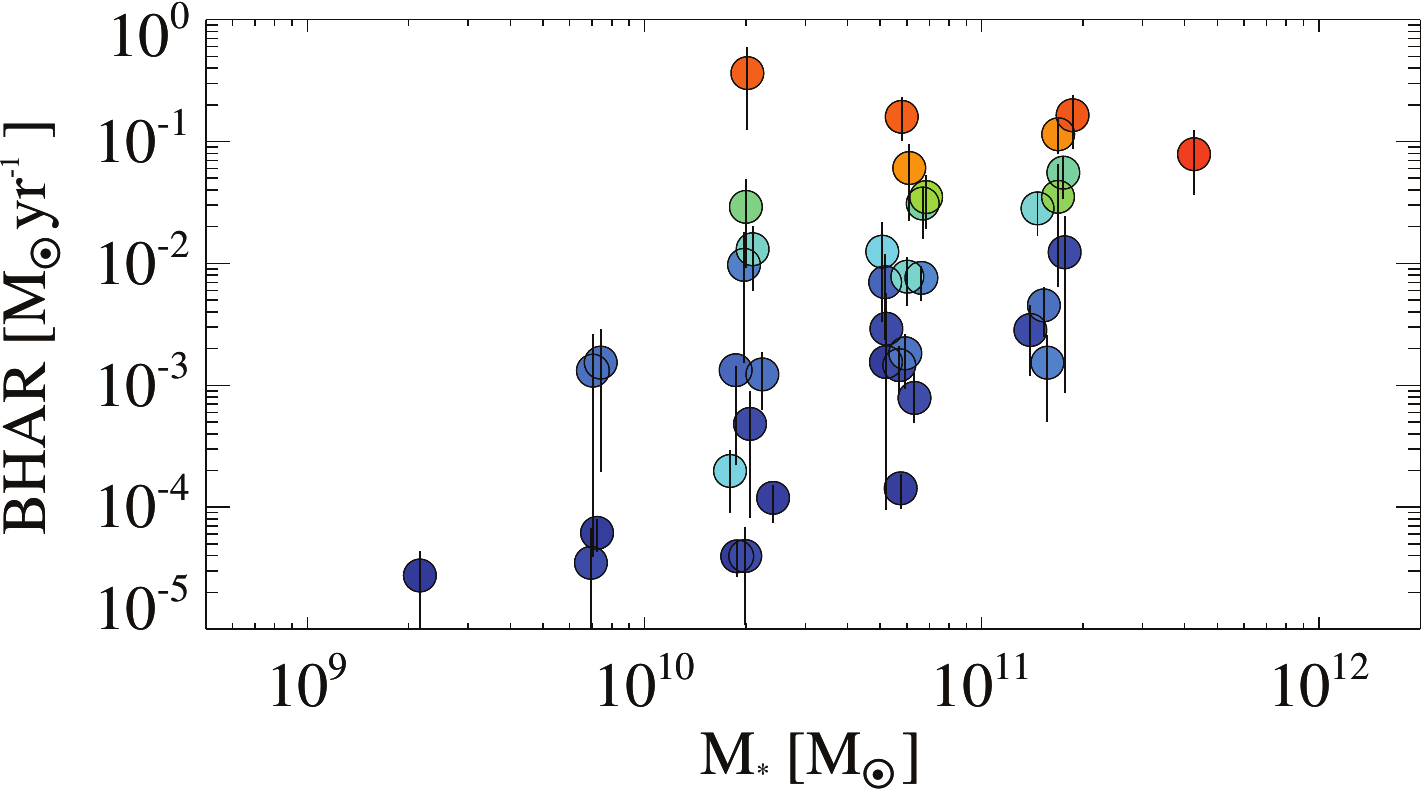}
\end{array}$
\end{center}
 \caption{\textit{Top panel}: distribution of bins including \textit{Herschel} galaxies above the $>$80 per cent completeness limit, as a function of BHAR, SFR and M$_*$ at all redshifts. The \textit{x}, {y} axes represent the SFR--M$_*$ plane, while the $z$-axis sets the colour-coded BHAR. Note that the three-dimensional space is logarithmic. \textit{Middle panel}: previous plot projected on the SFR--$\langle \dot{M}_{\rm bh} \rangle$ plane, with colour-coded redshift. \textit{Bottom panel}: projection on the M$_*$--$\langle \dot{M}_{\rm bh} \rangle$ plane, with colour-coded redshift. }
   \label{fig:3d_fit}
\end{figure}

After analysing redshift bins separately, we now merge results over the full redshift range covered by our study. The top panel of Fig. \ref{fig:3d_fit} shows the average BHAR (combining X-ray detected and stacked sources) in bins of SFR and M$_*$ from all our redshift slices. We have restricted ourselves here to the subsample with $>$80 per cent completeness in SFR. The central and bottom panels show the projections of the previous plot on the SFR--$\langle \dot{M}_{\rm bh} \rangle $ plane and M$_*$--$\langle \dot{M}_{\rm bh} \rangle $ plane, respectively, with a redshift colour-coding. 

In these panels, the data from different redshifts better connect into a tighter and more consistent sequence for BHAR=f(SFR) than for BHAR=f(M$_*$). This is in support of a primary dependence of the BHAR on SFR, though noting the less pronounced difference seen above within individual redshift slices.

\subsection{Comparison with previous studies}

We compare the relationships between AGN accretion and galaxy properties obtained from our analysis with those found by \citeauthor{Mullaney+12b} (\citeyear{Mullaney+12b}, M12 hereafter) and \citeauthor{Chen+13} (\citeyear{Chen+13}, C13 hereafter). M12 analysed a sample of star-forming galaxies in the GOODS-S using BzK and 24$~\mu$m selection. The authors claimed a linear relation at z$\sim$2 between $\langle \dot{M}_{\rm bh} \rangle $ and M$_*$, while Fig. \ref{fig:agn_m*} shows a weak trend. We checked whether the lack of trend in our data may be due to our sample selection. We match the selection criteria adopted by M12, by including main-sequence galaxies only (i.e. sSFR $\leq$ 4$\times$sSFR$_{\rm ms}$), and splitting the sample at 1.5$<$z$<$2.5 only in bins of M$_*$. In addition, we used a constant bolometric correction of 22.4 (from \citealt{Vasudevan+07}) to derive the average BHAR. We found that $\langle \dot{M}_{\rm bh} \rangle \propto$ M$_*^{0.86\pm0.08}$, while M12 derived $\langle \dot{M}
_{\rm bh} \rangle \propto$ M$_*^{1.05\pm0.36}$. However, this nearly linear relation vanishes as we split the 
sample as a function of both SFR and M$_*$: keeping the same M12-like sample and changing only the grid configuration, the resulting best-fit relation with M$_*$ is $\langle \dot{M}_{\rm bh} \rangle \propto$ M$_*^{0.45\pm0.25}$.\footnote{This trend is similar to what would be obtained in Fig. \ref{fig:agn_offset} (highest redshift bin) by removing the two bins corresponding to off-sequence galaxies (black squares).} 

To study the simultaneous dependence of $\langle \dot{M}_{\rm bh} \rangle $ on both SFR and M$_*$, we performed a multiple-correlation fitting in the $\log$ space, as a function of $\langle \dot{M}_{\rm bh} \rangle$, SFR and M$_*$ at the same time. We used a M12-like sample, but splitting it as a function of both SFR and M$_*$. We found that $\langle \dot{M}_{\rm bh} \rangle \propto \rm SFR^{1.04\pm0.29} \cdot M_*^{0.13\pm0.18}$. These slopes may suggest a ``pure'' SFR-driven relation.  

In summary, differences between our findings and M12 are due to the different binning and analysis: for an MS-only mass-binned sample where M$_*$ and SFR are degenerate because of the MS relation, we reproduce their close to linear relation of M$_*$ and BHAR, while covering the SFR--M$_*$ plane in both dimensions allows to separate the two variables and emphasizes the role of SFR.

Rodighiero et al. (2014, in preparation) analyse a mass-complete sample of galaxies at z$\sim$2 in the COSMOS field. We refer the reader to their work for a detailed comparison with results presented by M12.

We made also a comparison with results from C13, who measured a nearly linear trend between $\langle \dot{M}_{\rm bh} \rangle $ and SFR at 0.25$<$z$<$0.80, splitting their \textit{Herschel} sample in bins of SFR only. C13 found $\log$ BHAR = (-3.72$\pm$0.52) + (1.05$\pm$0.33) $\log$ SFR (both in M$_{\odot}~$yr$^{-1}$). By adjusting our grid configuration in bins of SFR to match their analysis, our fitting routine gives consistent coefficients, with BHAR = (-3.65$\pm$0.12) + (1.18$\pm$0.11) $\log$ SFR. We note that a similar trend is found even in case we split the sample in bins of SFR and M$_*$.

\section{Discussion} \label{discussion}

In Section \ref{correlations}, we described the relationship between BHAR and SFR, M$_*$ and offset from the main-sequence, over a wide range of galaxy properties and cosmic epochs, from z$\sim$0 up to z=2.5. In this Section we discuss and interpret our findings in the context of current AGN/galaxy evolutionary scenarios.

Mapping the average AGN accretion along both SFR and M$_*$ allows us to separate the degeneracy between the two parameters, shown by the presence of the MS relation since z$\sim$2. In Section \ref{corr_sfr_m*} we found that the average BHAR depends on both SFR and M$_*$ with similar significance levels. A more clear correlation with SFR is observed at z$>$0.8 and becomes more significant at z$\sim$2. This finding may support the SFR as the original driver of the correlation with average AGN accretion. 

We relate the trends of BHAR with SFR, M$_*$ and redshift to gas content and the redshift evolution of the gas fraction: 
\begin{equation}
f_{\rm gas} = \frac{M_{\rm gas}}{M_{\rm gas} + M_{*}} 
\end{equation}
which represents the ratio between total (i.e. molecular + neutral) gas mass and total baryonic mass of a galaxy. Several studies have investigated the evolution of $f_{\rm gas}$ in main-sequence galaxies from the local Universe to z$\sim$1 (\citealt{Leroy+08}; \citealt{Geach+11}; \citeauthor{Saintonge+11a} \citeyear{Saintonge+11a}, \citeyear{Saintonge+11b}, \citeyear{Saintonge+12}), at z$\sim$1.5--2 (\citealt{Daddi+10a}; \citealt{Tacconi+10}, \citeyear{Tacconi+13}), finding a strong ($f_{\rm gas} \propto$ (1+z)$^2$) evolution up to z$\sim$2 and a \textit{plateau} at higher redshift (z$\sim$3, \citealt{Magdis+13}). Direct CO observations presented by \citet{Daddi+10a} and \citeauthor{Tacconi+10} (\citeyear{Tacconi+10}, \citeyear{Tacconi+13}) provided first empirical evidence for the existence of very large gas fractions in z$\sim$1--2 main-sequence galaxies, with mass in gas even larger than the mass in stars (i.e. $f_{\rm gas} >$0.5). These gas rich systems have also higher SFRs (\citealt{Daddi+10b}; \
citealt{Genzel+10}), as 
expected from the Schmidt--Kennicutt relation (\citealt{Schmidt59}; \citealt{Kennicutt98}). This gas-dominated phase in z$\sim$1--2 main-sequence galaxies is also reflected in a difference in morphology. Indeed, while local main-sequence galaxies are preferentially regular disks, their z$\sim$2 analogs show a larger fraction of irregular morphologies and/or clumpy disks (e.g. \citealt{Elmegreen+07}; \citealt{Foerster-Schreiber+09}; \citealt{Kocevski+12}), which make them potentially more efficient in funnelling the cold gas inward onto the SMBH (\citealt{Bournaud+11}). 

In the light of the above-mentioned literature, it is justified to expect that the relationship between average BHAR and SFR becomes stronger with increasing redshift, where the fraction of actively star-forming gas becomes larger. Moreover, given that z$\sim$1--2 galaxies are truly gas rich, especially those at lower M$_*$ (e.g. \citealt{Santini+14}), it is also reasonable to assume that in these systems the fraction of \textit{primordial} gas (i.e. not yet converted into stars, therefore not causally linked to the galaxy stellar mass) is larger compared to local galaxies and to higher stellar mass galaxies at the same epoch. This may explain the weaker correlation of BHAR with M$_*$ that we found at higher redshift.

In line with these arguments, \citet{Vito+14} found that AGN hosts are significantly more gas rich than inactive galaxies, at a given M$_*$ and redshift, suggesting that the probability that a SMBH is active is strongly connected to the amount of cold gas supply. This supports M$_{\rm gas}$ as the key ingredient to explain the mutual evolution of star formation and AGN accretion activity (e.g. \citealt{Santini+12}; \citealt{Mullaney+12b}; \citealt{Rosario+13}; local AGN review by \citealt{Heckman&Best14}). However, the link between BHAR and SFR that we observe, suggests that AGN accretion is more closely related to the amount of cold \textit{star-forming} gas (M$_{\rm gas, SF}$), rather than to the total M$_{\rm gas}$. Though in this work we are not able to distinguish between nuclear and global star formation, we note that if AGN accretion were tracing global cold gas mass, the known strong differences in star-forming efficiency (or in ``depletion time'') between normal galaxies and luminous starbursts (e.g.
 \citealt{Solomon+97}; \citealt{Saintonge+12}) would predict a decreasing trend in BHAR/SFR with MS offset that we do not observe (section \ref{corr_offset}). 

As argued in Section \ref{corr_sfr_m*}, our analysis may support the SFR as the original driver of the correlation with BHAR, though not well discernible from M$_*$ within individual redshift slices. By combining all redshift bins, we are able to increase the statistics and infer more general trends between AGN accretion and galaxy properties. Fig. \ref{fig:3d_fit} shows the sub-sample of our \textit{Herschel} sources with FIR flux larger than the 80 per cent completeness threshold (see Section \ref{corr_offset}). The middle and bottom panels in Fig. \ref{fig:3d_fit} show that the overall correlation of BHAR with SFR is much narrower and significant than with M$_*$. 

We proceed to a comparison between the BHAR-SFR relation and the volume-averaged cosmic star-formation history and SMBH accretion history. As discussed in \citeauthor{Madau&Dickinson14} (\citeyear{Madau&Dickinson14}, see their Fig. 15), both the global SFR density (SFRD) and black hole accretion rate density (BHAD) peak around z$\sim$2, declining towards the local Universe. Despite the differences seen in recent derivations of accretion histories (e.g. \citealt{Shankar+09}; \citealt{Aird+10}; \citealt{Delvecchio+14}; \citealt{Ueda+14}), all of them systematically support a slightly faster decay of the BHAD since z$\sim$2 down to z$\sim$0, compared to the cosmic evolution of SFH. By fitting the BHAD/SFRD ratio in the $\log$--$\log$ space at 0$<$z$<$2, assuming the SFH from \citet{Madau&Dickinson14} and an average BHAD evolution from the afore-mentioned literature, we obtain BHAD $\propto$ SFRD$^{\rm 1.4\pm0.2}$. This relation is consistent with the slope evolution that we found between black hole and SFRs for 
our IR-selected sample: BHAR $\propto$ SFR$^{1.6 \pm 0.1}$ (see Fig. \ref{fig:3d_fit}, middle panel), supporting a scenario where the SMBH growth since z$\sim$2 follows a faster fading compared to their host galaxies. This redshift evolution is parametrized as BHAR/SFR $\propto$ (1+z)$^{0.9\pm0.3}$ that we found through a linear regression fit in the $\log$--$\log$ space. By integrating our estimates of average BHAR and SFR (see middle panel of Fig. \ref{fig:3d_fit}) over cosmic time yields to $\left( \int_{z=2}^{z=0}{\rm BHAR(z')~dz'} \right) $/$\left( \int_{z=2}^{z=0}{\rm SFR(z')~dz'} \right)$  $\approx$ 1/(3000${\pm 1500}$), marginally consistent with the prediction from the local M$_{\rm bh}$--M$_{\rm bulge}$ relation (around 1/1000, e.g. \citealt{Magorrian+98}). This slight difference has been found and extensively argumented in various studies (e.g. \citealt{Jahnke+09}; \citealt{Rafferty+11}; M12; C13). One possible factor is that our SFRs for main-sequence galaxies refer to the entire galaxy, 
including a 
large disk contribution, while only stars ultimately ending in the bulge should be counted in comparisons to the local M$_{\rm bh}$--M$_{\rm bulge}$ relation. Second, since in the local Universe the black hole accretion is less radiatively efficient compared to non-local AGN (\citealt{Merloni+08}), and the most powerful AGN are found in massive and passively evolving systems, a sample of radiatively efficient AGN might miss a not negligible fraction of the local BH accretion rate density, leading to a smaller BHAR/SFR ratio. Finally, heavily obscured AGN that are not detected by deep \textit{Chandra} surveys (e.g. \citealt{Donley+12}) may contribute to the cosmic AGN accretion history.

As described in Section \ref{corr_offset}, in none of our redshift slices did we find significant evidence for a correlation between BHAR and MS offset. While, at fixed M$_*$, outliers above the MS show enhanced BHAR (Fig. \ref{fig:agn_m*}), this is washed out in a larger sample such as ours, that includes MS and above-MS bins of similar SFR.

Recent studies based on deep \textit{Hubble Space Telescope} (HST) images, such as the Cosmic Assembly Near--infrared Deep Extragalactic Legacy Survey (CANDELS), allow thorough morphological analyses of star-forming galaxies at various redshifts. No difference in morphology between AGN hosts and ``inactive'' galaxies has been found at 0.7$<$z$<$3 (\citealt{Cisternas+11}; \citealt{Schawinski+11}; \citealt{Kocevski+11, Kocevski+12}; \citealt{Villforth+14}), which re-shapes the relevance of major mergers and supports a less violent picture, where secular processes (e.g clumpy and/or unstable disks, \citealt{Bournaud+12}) play a major role in triggering both AGN and star formation activity. Observations carried out with \textit{Herschel} (e.g. \citealt{Santini+12}; \citealt{Mullaney+12a}; \citealt{Rosario+13}) have shown that the FIR-based sSFRs of moderately luminous (L$\rm_X < $10$^{44}$ erg s$^{-1}$) X-ray selected AGN hosts up to z$\sim$3 are almost indistinguishable from those of inactive galaxies. 
According to these studies, it is plausible to find a non-trend between offset from the MS and AGN activity in our \textit{Herschel}-selected sample. The absence of a clear correlation is probably due to the fact that most of the luminous (L$_{\rm X}>$10$^{44}$ erg s$^{-1}$) AGN hosts are missed in our study due to the relatively small comoving volumes covered by our fields at low redshift.

A further question concerns how the BHAR/SFR ratio relates to major mergers. As noted, both morphology and total star formation of AGN hosts do not show evidence for a dominant role of major galaxy mergers in AGN triggering, of the type assumed by classical merger scenarios for the triggering of (luminous) AGN (\citealt{Sanders+88}; \citealt{Farrah+01}; \citealt{Springel+05}; \citealt{Hopkins+06}; \citealt{Sijacki+07}; \citealt{DiMatteo+08}). If outliers above the main sequence are mostly mergers, then questions on BHAR and BHAR/SFR of mergers link back to the discussion of BHAR at different positions with respect to the main sequence. Over our full M$_*$ and SFR range, we have found that MS offset is not a good predictor of absolute BHAR. However, at fixed M$_*$, the enhanced SFRs of above MS galaxies go along with enhanced BHAR, consistent with a BHAR/SFR ratio that does not change with position with respect to the main-sequence. While the merger enhances both SFR and BHAR, it does so (on average) at a 
ratio consistent with that of MS galaxies. 

In summary, our results suggest that accretion onto the black hole and star formation broadly and on average trace each other, irrespective of whether the galaxy is evolving steadily on the MS or bursting. This picture supports a causal connection between (radiatively efficient) AGN activity and amount of the cold star-forming gas. Because of the different spatial and variability scales of the two phenomena, this connection is apparent only in sample averages.

\section{Conclusions}

We analysed the \textit{average} BHAR in the SFR--M$_{*}$ plane at 0$<$z$\leq$2.5 and investigated for the first time the mutual relations between BHAR and SFR, M$_*$ and sSFR in about 8600 \textit{Herschel}-selected galaxies taken from the GOODS and COSMOS fields. The large statistics, together with the wealth of multi-wavelength data available in these fields, allow us to explore a wide variety of star-forming galaxies and to characterize their individual SEDs through broad-band SED-fitting decomposition. Average AGN bolometric luminosities and accretion rates have been derived from \textit{Chandra} X-ray observations, for both X-ray detected and undetected sources, under reasonable assumptions and widely used scaling factors. 

Our main conclusions are as follows.

\begin{enumerate}

 \item The average BHAR inferred in \textit{Herschel}-selected galaxies shows positive evolution as a function of both SFR and M$_*$ at z$<$0.8, while at higher redshift our data establish with $>$2$\sigma$ significance the SFR as the best predictor of AGN accretion. This may suggest that the galaxy SFR is the original driver of the correlation with accretion onto the central black hole. 
 
 \item Given the relation between BHAR and SFR, we found that the BHAR/SFR ratio slightly evolves with redshift, in agreement with a faster decline of the cosmic black hole accretion history with respect to the star formation history from z$\sim$2 to recent epochs, as published in the recent literature (e.g. Fig. 15 of \citealt{Madau&Dickinson14}). This evolutionary trend also leads to a lower BHAR/SFR ratio, albeit with a large associated error, compared to the predictions of the local M$_{\rm bh}$--M$_{\rm bulge}$ relation.
 
 \item We compared the observed correlations between AGN accretion and key galaxy properties with those presented by M12 and C13. Our analysis suggests that the $\langle \dot{M}_{\rm bh} \rangle $--M$_*$ correlation at z$\sim$2 claimed by M12 is likely a consequence of the trend with SFR and of the main-sequence relation that holds between the two. 

 \item These evolutionary trends of BHAR are plausible in the context of current studies of the evolution of the cold gas content in galaxies, if BHAR is on average linked to the content of dense star-forming gas.

\end{enumerate}

\section*{Acknowledgments}

The authors are grateful to the referee, J. R. Mullaney, for his constructive report.

This paper uses data from \textit{Herschel}'s photometers PACS and SPIRE. PACS has been developed by a consortium of institutes led by MPE (Germany) and including: UVIE (Austria); KU Leuven, CSL, IMEC (Belgium); CEA, LAM (France); MPIA (Germany); INAF-IFSI/OAA/OAP/OAT, LENS, SISSA (Italy) and IAC (Spain). This development has been supported by the funding agencies BMVIT (Austria), ESA-PRODEX (Belgium), CEA/CNES (France), DLR (Germany), ASI/INAF (Italy), and CICYT/MCYT (Spain). SPIRE has been developed by a consortium of institutes led by Cardiff Univ. (UK) and including: Univ.Lethbridge (Canada); NAOC (China); CEA, LAM (France); IFSI, Univ. Padua (Italy); IAC (Spain); Stockholm Observatory (Sweden); Imperial College London, RAL, UCL-MSSL, UKATC, Univ. Sussex (UK); and Caltech, JPL, NHSC, Univ. Colorado (USA). This development has been supported by national funding agencies: CSA (Canada); NAOC (China); CEA, CNES, CNRS (France); ASI (Italy); MCINN (Spain); SNSB (Sweden); STFC, UKSA (UK); and NASA (USA).

Funding for SDSS-III has been provided by the Alfred P. Sloan Foundation, the Participating Institutions, the National Science Foundation, and the U.S. Department of Energy Office of Science. The SDSS-III web site is \url{http://www.sdss3.org/}. SDSS-III is managed by the Astrophysical Research Consortium for the Participating Institutions of the SDSS-III Collaboration including the University of Arizona, the Brazilian Participation Group, Brookhaven National Laboratory, Carnegie Mellon University, University of Florida, the French Participation Group, the German Participation Group, Harvard University, the Instituto de Astrofisica de Canarias, the Michigan State/Notre Dame/JINA Participation Group, Johns Hopkins University, Lawrence Berkeley National Laboratory, Max Planck Institute for Astrophysics, Max Planck Institute for Extraterrestrial Physics, New Mexico State University, New York University, Ohio State University, Pennsylvania State University, University of Portsmouth, Princeton University, the 
Spanish Participation Group, University of Tokyo, University of Utah, Vanderbilt University, University of Virginia, University of Washington, and Yale University. 

Funding for PRIMUS is provided by NSF (AST-0607701, AST-0908246, AST-0908442, AST-0908354) and NASA (Spitzer-1356708, 08-ADP08-0019, NNX09AC95G). 

This research has made use of software provided by the \textit{Chandra} X-ray Center (CXC) in the application package CIAO (v. 4.6).

ID is grateful to Jillian Scudder for useful comments, and to Simonetta Puccetti and Nico Cappelluti for providing with X-ray sensitivity maps in the COSMOS field.  
ID is also thankful to Takamitsu Miyaji for his kind support in using CSTACK, and warmly thanks Fabio Vito for helpful suggestions with X-ray tools and routines to compute intrinsic X-ray luminosities. 
MB and GL acknowledge support from the FP7 Career Integration Grant ``eEASy'' (CIG 321913).

\bibliographystyle{mn2e}
\bibliography{biblio}

\appendix
\section{Details on X-ray stacking} \label{appendix_A}

The stacking analysis allows to increase the signal-to-noise of individually non-detected sources by grouping them together and piling-up their photon counts. This technique returns an average net (i.e. background subtracted) count rate for a given input list. As mentioned in Section \ref{stacking}, we stacked individual \textit{Herschel}-selected galaxies not detected in X-rays. We masked all point-like sources listed in public X-ray catalogues (\citealt{Alexander+03} for GOODS-N, \citealt{Xue+11} for GOODS-S and \citealt{Elvis+09} for C-COSMOS). We have assumed a circular shape for each mask, whose radius is 50 per cent larger then the area enclosing 90 per cent of photon counts from the source. The size of the Point Spread Function (PSF) follows a radial (and energy-dependent) profile in the \textit{Chandra} Deep Fields, while in COSMOS it shows fluctuations in between overlapping pointings (see \citealt{Elvis+09}). We accounted for that and ensured that the mean background level, after masking all X-ray 
sources, was fully consistent in each field with that presented in the corresponding reference papers.

To optimize the final signal-to-noise ratio (SNR), we have taken a fixed source aperture (i.e. 2 arcsec radius) centered on each optical position. The photon counts have been corrected for off-axis angle and observed frequency\footnote{This correction has been made using the \textit{psf} module (\url{http://cxc.harvard.edu/ciao/ahelp/psf.html}) from the \textit{Chandra} Interactive Analysis of Observations (CIAO) package. }.

Each background region is an annulus centered on the corresponding optical position, with inner radius placed 5 pixels (1 pixel = 0.492 arcsec) apart from the PSF size enclosing 90 per cent source counts. The outer radius is 5 pixels apart from the inner one. This avoids contamination from the PSF of the stacked source.

We subtracted the background counts, rescaled to the source area, from the photon counts collected within the source region, to get the net source counts. The exposure time is taken from energy-dependent time-maps, which return for each (x,y) position in the sky the effective exposure time (i.e. corrected for vignetting, bad pixels, dithering and including also a spatially dependent quantum efficiency). Finally, the average count rate is computed by summing (over all stacked sources) all net source counts and dividing by the total effective exposure time.

\section{Statistics in the SFR--M$_*$ plane} \label{appendix_B}

In Table \ref{tab:data} we show the values of BHAR and its uncertainties, as a function of SFR, M$_*$ and redshift, in order to quantify the colour-coded scale shown in Fig. \ref{fig:plane_agn}. In addition, we list the number of sources included in each bin (total numbers, X-ray detected and X-ray undetected). Note that bins of SFR change as a function of redshift, as they are arranged to follow the cosmic evolution of the main-sequence relation.

\begin{table*}
   \caption{List of average BHAR, as a function of SFR, M$_*$ and redshift, with related 1$\sigma$ uncertainties. Alongside of each redshift range, we report the median redshift value $\langle z \rangle $ and the sSFR$_{\rm ms}$ that follows the main-sequence at $\langle z \rangle $. In some bins only an upper limit (upp) or a lower limit (low) is available. Numbers between parentheses in each bin represent: total number of \textit{Herschel} sources, X-ray detected and X-ray stacked, respectively. }
\begin{tabular}{lcccccc}
\hline
		    &               &                  &    0.01 $\leq$ z $<$ 0.25  &    $\langle z \rangle $ = 0.18        &     sSFR$_{\rm ms}$ = 1.27e-10 yr$^{-1}$               &                \\
  \hline
    SFR [M$_{\odot}$yr$^{-1}$] 	    &               &                  &   log (M$_*$/M$_{\odot}$)            &                 &                  &                 \\
		    &  9.00 -- 9.50 &  9.50 -- 10.00   & 10.00 -- 10.50         & 10.50 -- 11.00  &  11.00 -- 11.50  & 11.50 -- 12.00  \\
\hline
                  &               &                  &     &                 &                  &                \\ 
0.13 -- 0.40      & -5.00$^{+0.24}_{-0.26}$  &   &   &   &   &   \\
                   &  (25         4        21)           &             &                     &              &               &             \\
                  &               &                  &     &                 &                  &                \\ 
0.40 -- 1.27      &    -4.40$^{+0.35}_{-0.36}$ &  -4.56$^{+2.24}_{-0.27}$  &  -5.84 (low) &  -4.04$^{+0.16}_{-0.20}$  &   &   \\
                   &  (31         3        28)           &  (51         4        47)      &  (32         3        29)    &  (22         5        17)           &               &               \\
                  &               &                  &     &                 &                  &                \\                    
 1.27 -- 4.00      &   & -4.12$^{+0.09}_{-0.13}$ &  -4.43$^{+0.14}_{-0.20}$  &  -2.77$^{+0.32}_{-0.78}$ &   &   \\
                   &       &  (51         3        48)    &    (89         8        81)      &      (34         7        27)     &              &         \\
                  &               &                  &     &                 &                  &                \\                    
 4.00 -- 12.66     &   &  & -3.92$^{+0.15}_{-0.29}$  & -3.85$^{+0.17}_{-0.29}$  &   &   \\
                  &               &          &    (19         3        16)   &    (33         6        27)               &                  &                \\                    
		  &               &                  &     &                 &                  &                \\                                      
\hline

		    &               &                  &    0.25 $\leq$ z $<$ 0.50  &    $\langle z \rangle $ = 0.35        &     sSFR$_{\rm ms}$ = 1.79e-10 yr$^{-1}$             &                \\
  \hline
    SFR [M$_{\odot}$yr$^{-1}$] 	    &               &                  &   log (M$_*$/M$_{\odot}$)            &                 &                  &                 \\
		    &  9.00 -- 9.50 &  9.50 -- 10.00   & 10.00 -- 10.50         & 10.50 -- 11.00  &  11.00 -- 11.50  & 11.50 -- 12.00  \\
\hline
                  &               &                  &     &                 &                  &                \\ 
0.57 -- 1.79      &  & -4.21$^{+0.34}_{-0.48}$  & -5.65 (upp)  &   &   &   \\
                  &  & (36         2        34)  & (29         0        29)  &   &   &   \\
                   &  &   &   &   &   &   \\
1.79 -- 5.67       & -3.89$^{+0.45}_{-0.63}$ & -3.98$^{+0.48}_{-0.98}$  & -4.26$^{+0.20}_{-0.40}$  &  -2.50$^{+0.33}_{-0.64}$ &  -3.53$^{+0.20}_{-0.26}$ &   \\
                    &  (24         2        22) & (128         4       124)  & (287         9       278)  & (177        12       165)  &  (38         5        33) &   \\
                    &  &   &   &   &   &   \\
5.67 -- 17.94      &  &  -4.10 (upp) & -3.42$^{+0.42}_{-0.86}$  &  -2.71$^{+0.21}_{-0.32}$ & -2.62$^{+0.23}_{-0.38}$  &   \\
                    &  &  (15         2        13) & (104        10        94)  &  (208        21       187) & (59        10        49)  &   \\
                   &  &   &   &   &   &   \\
17.94 -- 56.72     &  &   &   &   -3.12$^{+0.25}_{-0.28}$ & -1.91$^{+0.56}_{-1.50}$  &   \\
                   &  &   &   &  (31         7        24) & (16         4        12)  &   \\                    
		   &  &   &   &   &   &   \\
\hline

		    &               &                  &    0.50 $\leq$ z $<$ 0.80  &    $\langle z \rangle $ = 0.65        &     sSFR$_{\rm ms}$ = 3.13e-10 yr$^{-1}$        &                \\
  \hline
    SFR [M$_{\odot}$yr$^{-1}$] 	    &               &                  &   log (M$_*$/M$_{\odot}$)            &                 &                  &                 \\
		    &  9.00 -- 9.50 &  9.50 -- 10.00   & 10.00 -- 10.50         & 10.50 -- 11.00  &  11.00 -- 11.50  & 11.50 -- 12.00  \\
\hline
                  &               &                  &     &                 &                  &                \\ 
0.99 -- 3.13       &  & -4.59$^{+0.31}_{-0.94}$  &   &   &   &   \\
                   &  &  (15         1        14) &   &   &   &   \\
                   &  &   &   &   &   &   \\
3.13 -- 9.91        &  & -3.20$^{+0.43}_{-0.83}$  & -3.27$^{+0.05}_{-0.47}$  &  -2.51$^{+0.20}_{-0.30}$ &  -1.79$^{+0.45}_{-1.37}$ &   \\ 
                     &  & (78         4        74)  &  (171        13       158) & (141        11       130)  & (34         4        30)  &   \\
                     &  &   &   &   &   &   \\
9.91 -- 31.33      &  & -2.16$^{+0.16}_{-1.13}$  & -3.29$^{+0.16}_{-0.25}$  & -2.85$^{+0.15}_{-0.21}$  &  -1.98$^{+0.18}_{-0.38}$ &   \\
                    &  &  (80         8        72) & (300        16       284)  & (522        26       496)  & (168        28       140)  &   \\
                   &  &   &   &   &   &   \\
31.33 -- 99.08     &  &   & -2.07$^{+0.49}_{-0.94}$  & -2.07$^{+0.17}_{-0.35}$  &  -2.81$^{+0.37}_{-0.64}$ &   \\
                    &  &   & (20         2        18)  & (   82        18        64)  &  (53         7        46) &   \\
		    &  &   &   &   &   &   \\
\hline

\label{tab:data}
\end{tabular}

\end{table*}

\begin{table*}

\begin{tabular}{lcccccc}
\hline

&               &                  &    0.80 $\leq$ z $<$ 1.50  &       $\langle z \rangle $ = 1.00      &     sSFR$_{\rm ms}$ = 5.54e-10 yr$^{-1}$        &                \\
  \hline
    SFR [M$_{\odot}$yr$^{-1}$] 	    &               &                  &   log (M$_*$/M$_{\odot}$)            &                 &                  &                 \\
		    &  9.00 -- 9.50 &  9.50 -- 10.00   & 10.00 -- 10.50         & 10.50 -- 11.00  &  11.00 -- 11.50  & 11.50 -- 12.00  \\
\hline
                  &               &                  &     &                 &                  &                \\ 
5.54 -- 17.52       &  & -3.28$^{+0.16}_{-0.16}$  &  -2.77$^{+0.21}_{-0.27}$ &  -2.15$^{+0.15}_{-0.22}$ &  -1.91$^{+0.33}_{-0.74}$ &   \\
                   &  &  (56         3        53) &  (154         7       147) &  (202        15       187) &  (69         8        61) &   \\
                   &  &   &   &   &   &   \\
17.52 -- 55.40        &  &  -2.30$^{+0.53}_{-1.01}$ & -2.38$^{+0.15}_{-0.23}$  & -2.16$^{+0.11}_{-0.13}$  &  -1.77$^{+0.11}_{-0.14}$ &   \\ 
                     &  & (71         5        66)  &  (420        18       402) & (985        65       920)  & (501        55       446)  &   \\
                     &  &   &   &   &   &   \\
55.40 -- 175.18      &  &   &  -1.40$^{+0.32}_{-0.50}$  &  -1.38$^{+0.20}_{-0.36}$ &  -1.36$^{+0.19}_{-0.29}$  & -1.72$^{+0.33}_{-0.71}$  \\
                    &  &   & (91        11        80)  &  (339        37       302) & (296        42       254)  &  (19         1        18) \\
                   &  &   &   &   &   &   \\
175.18 -- 553.98     &  &   &   & -1.46$^{+0.34}_{-0.38}$  & -1.41$^{+0.35}_{-0.95}$  &   \\
                    &  &   &   & (18         5        13)  &  (19         3        16) &   \\
		    &  &   &   &   &   &   \\
\hline

		    &               &                  &    1.50 $\leq$ z $\leq$ 2.50  &        $\langle z \rangle $ = 1.91      &     sSFR$_{\rm ms}$ = 1.79e-9 yr$^{-1}$           &                \\
  \hline
  SFR [M$_{\odot}$yr$^{-1}$]	    &               &                  &   log (M$_*$/M$_{\odot}$)            &                 &                  &                 \\
		    &  9.00 -- 9.50 &  9.50 -- 10.00   & 10.00 -- 10.50         & 10.50 -- 11.00  &  11.00 -- 11.50  & 11.50 -- 12.00  \\
\hline
                  &               &                  &     &                 &                  &                \\ 
17.92 -- 56.68       &  &  -1.03$^{+0.20}_{-1.29}$  & -2.44$^{+0.40}_{-0.61}$  & -1.74$^{+0.16}_{-0.24}$  &   -1.38$^{+0.40}_{-0.95}$ &   \\
                   &  &  (19         3        16) &  (47         2        45) & (103        11        92)  & (35         3        32)  &   \\
                   &  &   &   &   &   &   \\
56.68 -- 179.24        &  & -0.53$^{+0.34}_{-0.36}$  &  -0.82$^{+0.27}_{-0.39}$  &  -1.50$^{+0.18}_{-0.29}$  & -1.13$^{+0.12}_{-0.15}$  &  -0.57$^{+0.20}_{-0.25}$ \\ 
                     &  &  (16         4        12) &  (125        20       105) &  (230        19       211) &  (272        43       229) &  (38        12        26) \\
                     &  &   &   &   &   &   \\
179.24 -- 566.80      &  &   & -0.49$^{+0.49}_{-0.81}$  &   -1.01$^{+0.27}_{-0.39}$ & -0.63$^{+0.25}_{-0.49}$  & -0.87$^{+0.21}_{-0.32}$  \\
                    &  &   & (22         6        16)  &  (75        14        61) & (128        24       104)  &  (34        10        24) \\
                   &  &   &   &   &   &   \\
\hline

\end{tabular}

\end{table*}

\label{lastpage}

\end{document}